%% file: main_arxiv.tex
\documentclass[a4paper]{article}
\usepackage{latexsym,amsthm,amsmath,amssymb}

\newtheorem{observation}{Observation}
\newtheorem{definition}{Definition}

\theoremstyle{plain}
\newtheorem{rrule}{Reduction Rule}[section]

\newtheorem{lemma}{Lemma}
\newtheorem{theorem}{Theorem}

\usepackage[T1]{fontenc}
\usepackage{bbm}
\usepackage{mathtools}
\usepackage{amsfonts}
\usepackage{pifont}
\usepackage{url}
\usepackage{multirow}
\usepackage{multicol}
\usepackage{fancybox}
\usepackage{colortbl}
\usepackage{soul}
\usepackage{color}
\usepackage{tikz}
\usepackage{tkz-berge}
\usepackage{rotating}
\usepackage{pdflscape}
\usepackage{afterpage}
\usepackage{comment}
\usepackage{etoolbox}
\usepackage{environ}
\usepackage{dirtytalk}
\usepackage{authblk}

\usepackage{cite}
\usepackage{hyperref}

\usepackage{algorithm}
\usepackage{algorithmic}\usetikzlibrary{decorations.pathmorphing}

\usetikzlibrary{decorations,arrows,shapes}
\newcommand{\inners}{1.2pt}
\newcommand{\outers}{1pt}
\newcommand{\angled}[1]{\left\langle{#1}\right\rangle}

\usepackage{complexity}
\newclass{\Hard}{hard}
\newclass{\pNP}{paraNP}
\newclass{\Hness}{hardness}
\newcommand{\NPH}{\NP\text{-}\Hard}

\newclass{\Complete}{complete}
\newclass{\Total}{Total}
\newclass{\Delay}{Delay}
\newclass{\Cness}{completeness}
\newclass{\PPT}{PPT}
\newcommand{\NPc}{\NP-\Complete}

\newcommand{\TFPT}{\Total\FPT}
\newcommand{\DFPT}{\Delay\FPT}
\newfunc{\bits}{bits}
\newfunc{\Ext}{Ext}
\newfunc{\Sol}{Sol}
\newfunc{\YES}{YES}
\newfunc{\NOi}{NO}
\newfunc{\tw}{tw}
\newfunc{\dcc}{dcc}
\newfunc{\dc}{dc}
\newfunc{\roott}{root}
\newcommand{\pname}[1]{\textsc{#1}}
\newcommand{\WH}[1]{\W[#1]\text{-}\Hard}
\newcommand{\WHness}[1]{\W[#1]\text{-}\Hness}

\newfunc{\dist}{dist}
\newfunc{\diam}{diam}
\newfunc{\border}{border}
\newfunc{\dgr}{deg}

\newfunc{\Rep}{Rep}
\newfunc{\DPx}{DP}

\newfunc{\leavesx}{leaves}

\newfunc{\opt}{opt}
\newfunc{\rmcx}{rmc}

\newfunc{\ins}{ins}
\newfunc{\shift}{shift}
\newfunc{\glue}{glue}
\newfunc{\proj}{proj}
\newfunc{\joinf}{join}
\newfunc{\up}{up}
\newfunc{\down}{down}

\newcommand{\bigO}[1]{\mathcal{O}\!\left(#1\right)}
\newcommand{\bigOs}[1]{\mathcal{O}^*\!\left(#1\right)}

\newcommand{\probl}[3]{
  \begin{flushleft}
    \fbox{
      \begin{minipage}{0.97\linewidth}
        \noindent {\sc #1}\\
        {\bf Instance:} #2\\
        {\bf Question:} #3
      \end{minipage}}
    \medskip
  \end{flushleft}
}

\newcommand{\problenum}[3]{
  \begin{flushleft}
    \fbox{
      \begin{minipage}{0.97\linewidth}
        \noindent {\sc #1}\\
        {\bf Instance:} #2\\
        {\bf Enumerate:} #3
      \end{minipage}}
    \medskip
  \end{flushleft}
}

\NewEnviron{sproof}[1] {
    \begin{proof}[Proof of safeness of Rule~#1]\BODY\end{proof}
}

\title{Matching (Multi)Cut: Algorithms, Complexity, and Enumeration}
\date{}

\author[1]{Guilherme C. M. Gomes}
\author[1]{Emanuel Juliano}
\author[1]{Gabriel Martins}
\author[1]{Vinicius F. dos Santos}

\affil[1]{Departamento de Ci\^encia da Computa\c{c}\~{a}o, Universidade Federal de Minas Gerais -- Belo Horizonte, Brazil}

\begin{document}

\maketitle

\begin{abstract}
    A matching cut of a graph is a partition of its vertex set in two such that no vertex has more than one neighbor across the cut. The Matching Cut problem asks if a graph has a matching cut. This problem, and its generalization d-cut, has drawn considerable attention of the algorithms and complexity community in the last decade, becoming a canonical example for parameterized enumeration algorithms and kernelization. In this paper, we introduce and study a generalization of Matching Cut, which we have named Matching Multicut: can we partition the vertex set of a graph in at least $\ell$ parts such that no vertex has more than one neighbor outside its part?
    We investigate this question in several settings.
    We start by showing that, contrary to Matching Cut, it is NP-hard on cubic graphs but that, when $\ell$ is a parameter, it admits a quasi-linear kernel. We also show an $\mathcal{O}(\ell^{\frac{n}{2}})$ time exact exponential algorithm for general graphs and a $2^{\mathcal{O}(t\log t)}n^{\mathcal{O}(1)}$ time algorithm for graphs of treewidth at most $t$.
    We then parameterized enumeration aspects of matching multicuts. First, we generalize the quadratic kernel of Golovach et. al for Enum Matching Cut parameterized by vertex cover, then use it to design a quadratic kernel for Enum Matching (Multi)cut parameterized by vertex-deletion distance to co-cluster. Our final contributions are on the vertex-deletion distance to cluster parameterization, where we show an FPT-delay algorithm for Enum Matching Multicut but that no polynomial kernel exists unless NP $\subseteq$ coNP/poly; we highlight that we have no such lower bound for Enum Matching Cut and consider it our main open question.
\end{abstract}

\input{intro}
\input{prelim}

\input{np_cubic}
\input{exact}
\input{treewidth}
\section{Matching Multicut Enumeration}
\input{enum_vc}

\input{enum_dcc}

\input{enum_dc}
\input{kernel_lb}
\input{final}

\section*{Acknowledgements}

This work was partially funded by Coordenação de Aperfeiçoamento de Pessoal de Nível Superior -- Brasil (CAPES) -- Código de Financiamento 001, by Conselho Nacional de Desenvolvimento Científico e Tecnológico (CNPq), grants 312069/2021-9, 406036/2021-7 and 404479/2023-5 by Fundação de Amparo à Pesquisa do Estado de Minas Gerais (FAPEMIG), grant APQ-01707-21.

\newpage
\bibliographystyle{plainurl}
\bibliography{main_arxiv}


\end{document}

%% file: intro.tex
\section{Introduction}

A matching $M$ in a graph $G$ is a subset of the edges of $G$ such that no vertex is the endpoint of more than one edge in $M$.
Matchings are one of the most fundamental concepts in graph theory, with whole books dedicated to them~\cite{lovasz2009matching,lucchesi2024perfect}. A cut of a graph $G$ is a partition of its vertex set into two non-empty sets and we say that the set of edges between them is an edge cut. A matching cut is a edge cut that is also a matching.

Not all graphs admit a matching cut, and graphs admitting such kind of cuts were first considered by Graham~\cite{graham1970primitive}, which called them decomposable graphs, to solve a problem in number theory. Other applications include fault-tolerant networks~\cite{farley1982networks}, multiplexing networks~\cite{araujo2012good} and graph drawing~\cite{patrignani2001complexity}.
The problem of recognizing graphs that do admit a matching cut, called \pname{Matching Cut}, was studied by Chvátal~\cite{chvatal1984recognizing}, who proved that the problem is \NPc\ even restricted to graphs of maximum degree four, while polynomial-time solvable in graphs of maximum degree three. The problem was reintroduced under the current terminology in~\cite{patrignani2001complexity} and, since then, it has been attracting much attention of the algorithms community. It has also been shown to remain \NPc\ for several graph classes such as bipartite graphs of bounded degree~\cite{randerath2003stable}, planar graphs of bounded degree or bounded girth~\cite{bonsma2009complexity} and $P_t$-free graphs (for large enough $t$)~\cite{feghali2023note}. On the positive side, tractable cases include $H$-free graphs, i.e. graphs without an induced subgraph isomorphic to $H$, for some $H$, including $P_6$, the path on 6 vertices~\cite{lucke2022complexity}. For a more comprehensive overview and recent developments, we refer to~\cite{chen2021matching,lucke2023finding}.


\pname{Matching Cut} has also been studied from the parameterized perspective, with the minimum number of edges crossing the cut $k$ being used as the natural parameter for this problem.
The first parameterized algorithm for $k$ was given by Marx et al. in~\cite{treewidth_reduction}; they tackled the \pname{Stable Cutset} problem using the treewidth reduction machinery and Courcelle's theorem, which yielded a very large dependency on $k$.
We remark that \pname{Matching Cut} on $G$ is equivalent to finding a separator that is an independent set in the line graph of $G$.
Using the compact tree decomposition framework of Cygan et al.~\cite{compact_td}, Aravind and Saxena~\cite{fpt_crossing_mc_dcut} developed a $2^{\bigO{k \log k}}$ time algorithm for \pname{Matching Cut}.
Komusiewicz et al.~\cite{matching_cut_ipec} presented a quadratic kernel for the vertex-deletion distance to cluster parameterization, as well as single exponential time \FPT\ algorithms for this parameterization and for vertex-deletion distance to co-cluster; on the other hand, they gave a kernelization lower bound for the combined parameterization of treewidth plus the number of edges in the cut.
Aravind et al.~\cite{matching_cut_structural} presented \FPT\ algorithms for neighborhood diversity, twin-cover and treewidth for \pname{Matching Cut}; the latter had its running time improved by Gomes and Sau in~\cite{dcut_gg}.

One area in which matching cuts have drawn particular attention is in parameterized enumeration.
Under this framework, our goal is to list all feasible solutions to a problem, e.g. all matching cuts of an input graph.
Parameterized algorithms that do so are classified in two families: \TFPT\ --  where all solutions can be listed in \FPT\ time -- and \DFPT\ -- where the delay between outputting two solutions, i.e. the time between these outputs, is at most \FPT.
Based on the foundational work of Creignou et al.~\cite{creignou2017enum}, Golovach et al.~\cite{golovach2022refined} defined the kernelization analogues of \TFPT\ and \DFPT.
Also in~\cite{golovach2022refined}, the authors developed several enumeration and kernelization algorithms for \pname{Enum Matching Cut} under the vertex cover, neighborhood diversity, modular width, and clique partition number parameterizations.
They also studied the enumeration of minimal and maximal matching cuts in the form of the \pname{Enum Minimal MC} and \pname{Enum Maximal MC} problems under some of the aforementioned parameterizations.

Similar problems to \pname{Matching Cut}, as well as minimization and maximization questions~\cite{lucke2023dichotomies}, have also been considered. Their hardness follow directly from the problem definition. Another related problem, \pname{Perfect Matching Cut}, asks for the existence of a perfect matching that is also a matching cut. Although its hardness does not follow directly from \pname{Matching Cut} the problem is also \NPc~\cite{heggernes1998partitioning}. The recent survey by Le et al.~\cite{le2023maximizing} revisit and compare results on these variations.
Some problems, however, can be seen as direct generalizations of \pname{Matching Cut}.
In the $d$-\pname{Cut} problem, the goal is to partition the vertex set into two sets such that each vertex has at most $d$ neighbors in the opposite set of the partition. 
Introduced in~\cite{dcut_gg}, $d$-\pname{Cut} has been shown to be \NPc\ for $(2d+2)$-regular graphs and it has been shown to admit \FPT\ algorithms under several parameters such as the maximum number of edges crossing the cut~\cite{fpt_crossing_mc_dcut}, treewidth, vertex-deletion distance to cluster, and vertex-deletion distance to co-cluster.
When $d=1$ the problem is exactly \pname{Matching Cut}. However, many cases that are tractable for $d=1$ have been shown to become hard for $d$-\pname{Cut}~\cite{lucke2024finding}.
The other related problem arises in the context of graph convexity. To our purposes, a convexity is a family $\mathcal{C}$ of subsets of a finite ground set $X$ such that $X, \emptyset \in \mathcal{C}$ and $\mathcal{C}$ is closed for intersection. Many graph convexities have been considered in the literature~\cite{harary1981convexity, dourado2010complexity, araujo2018convexity, centeno2011irreversible}, most of them motivated by families of paths. In this context, a subset $S$ of vertices is convex if all paths of a given type between vertices of $S$ contain only vertices of $S$. The most well-studied paths in the literature are shortest paths, induced paths and $P_3$, the paths on three vertices. One of the problems studied in the graph convexity setting is the partition of the vertex set of a graph into convex sets. Note that, in the $P_3$-convexity, this is equivalent to partition vertices in such a way that two vertices in a set $S$ have no common neighbor outside $S$. Hence, partitioning into two $P_3$-convex sets is equivalent to \pname{Matching Cut}. The more general case has also been considered in~\cite{centeno2010convex, gonzalez2020covering}.

\smallskip
\noindent \textbf{Our contributions.} In this work we introduce the \pname{Matching Multicut} problem, a novel generalization of  \pname{Matching Cut}. A \textit{matching multicut on $\ell$ parts} of a graph $G$ is a partition of its vertex set in $\{A_1, \dots, A_{\ell}\}$ such that each vertex in $A_i$ has at most one neighbor outside of $A_i$. Note that this is quite different from a partition into $P_3$-convex sets; in the latter, a vertex $v \in A_i$ may have one neighbor in \textit{each} other $A_j$, while in the former, $v$ may have one neighbor in $\bigcup_{j \neq i} A_j$.
Formally, we study the following problem:

\probl{Matching Multicut}{A graph $G$ and an integer $\ell$.}{Does $G$ admit a matching multicut on at least $\ell$ parts?}

We explore the complexity landscape of \pname{Matching Multicut} under several settings that were previously considered for \pname{Matching Cut}. Since the case $\ell=2$ is exactly \pname{Matching Cut}, the problem is trivially \NPH. It is also trivially \WH{1}\ for the natural parameter $\ell$. We study its complexity for cubic graphs, exact exponential algorithms, structural parameterizations as well parameterized enumeration aspects of matching multicuts.

Contrary to the classic result of Chvátal showing the polynomial-time solvability of \pname{Matching Cut}~\cite{chvatal1984recognizing} for cubic graphs, we show that \pname{Matching Multicut} is \NPH\ even restricted to those graphs. On the other hand, we show that the problem becomes fixed parameter tractable when parameterized by $\ell$. Indeed, we show that the problem admits a quasi-linear kernel under this parameterization. We also show that the problem is \FPT\ parameterized by treewidth.
From the definition of the problem, there is a trivial  $\ell^nn^{\bigO{1}}$ time algorithm for \pname{Matching Multicut}
by just enumerating all possible (ordered) partitions of $V(G)$.
We improve this by showing that the problem can be solved in $\alpha_\ell^nn^{\bigO{1}}$ time, with $\alpha_\ell \leq \sqrt{\ell}$ for a graph on $n$ vertices.
Finally, we turn our attention to the enumeration of matching multicuts in the form of the \pname{Enum Matching Multicut} problem.

\problenum{Enum Matching Multicut}{A graph $G$ and an integer $\ell$.}{All matching multicuts of $G$ on at least $\ell$ parts}

Our first results in this direction are a PDE kernel under vertex cover and a PDE kernel under vertex-deletion distance to co-cluster.
Afterwards, we present a \DFPT\ algorithm for enumerating matching multicuts of a graph parameterized by the vertex-deletion distance to cluster.
Our final result is a negative one.
We show that, although \pname{Enum Matching Multicut} is in \DFPT\ for the vertex-deletion distance to cluster parameter, we show that \pname{Matching Multicut} does not admit a polynomial kernel under the joint parameterization of distance to cluster, maximum cluster size and the number of parts of the cut, unless $\NP \subseteq \coNP/\poly$. To prove this result, we show that \pname{Set Packing} has no polynomial kernel parameterized by the size of the ground set, which could be of independent interest. To the best of our knowledge, although expected, this result has not been explicitly stated before.

%% file: prelim.tex
\subsection{Preliminaries}

We denote $\{1, 2, \dots, n\}$ by $[n]$.
We say that a monotonically non-decreasing function $f$ is \textit{quasi-linear} if $f \in \mathcal{n \log^c n}$ for some constant $c$.
We use standard graph-theoretic notation, and we consider simple undirected graphs without loops or multiple edges; see~\cite{murty} for any undefined terminology. When the graph is clear from the context, the degree (that is, the number of neighbors) of a vertex $v$ is denoted by  $\deg(v)$, and the number of neighbors of a vertex $v$ in a set $A \subseteq V(G)$ and its neighborhood in it are denoted by $\deg_A(v)$ and $N_A(v)$; we also define $N(S) = \bigcup_{v \in S} N(v) \setminus S$.
The minimum degree and the maximum degree of a graph $G$ are denoted by $\delta(G)$ and $\Delta(G)$, respectively.
We say that $G$ is cubic if $\deg(v) = 3$ for all $v \in V(G)$ and that $G$ is subcubic if $\deg(v) \leq 3$.
A \textit{matching} $M$ of $G$ is a subset of edges of $G$ such that no vertex of $G$ is incident to more than one edge in $M$; for simplicity, we define $V(M) = \bigcup_{uv \in M} \{u,v\}$ and refer to it as the set of \textit{$M$-saturated vertices}.
The \textit{subgraph of $G$ induced by $X$} is defined as $G[X] = (X, \{uv \in E(G) \mid u,v \in X\})$.
The vertex-deletion distance to $\mathcal{G}$ is the size of a minimum cardinality set $U \subseteq V(G)$ such that $G \setminus U = G[V(G) \setminus U]$ belongs to class $\mathcal{G}$; in this case, $U$ is called the $\mathcal{G}$-modulator.
A graph $G$ is a \textit{cluster graph} if each connected component is a clique; $G$ is a co-cluster graph if its complement is a cluster graph. A \textit{vertex cover} of $G$ is a set of vertices incident to every edge of $G$.
A tree decomposition $(T, \{X_t\}_{t \in V(T)})$ of a connected graph $G$ is such that $T$ is a tree, $X_t \subseteq V(G)$ for all $t$ and: (i) for every $uv \in E(G)$ there is some $t \in T$ where $u,v \in X_t$ and (ii) the nodes of $T$ that contain $v \in V(G)$ form a subtree of $T$, for every $v$.
The sets $X_t$ are called the \textit{bags} of the decomposition, the \textit{width} of the decomposition is $\max_{t \in V(T)} X_t - 1$.
The \textit{treewidth} of $G$ is the size of a tree decomposition of $G$ of minimum width. For more on treewidth and, in particular, nice tree decompositions, we refer the reader to~\cite{cygan_parameterized}.

We refer the reader to~\cite{downey_fellows,cygan_parameterized} for basic background on parameterized complexity, and we recall here only some basic definitions.
A \emph{parameterized problem} is a tuple $(L, \kappa)$ where $L \subseteq \Sigma^*$ is a language and $\kappa: \Sigma^* \mapsto \mathbb{N}$ is a parameterization.  For an instance $I=(x,k) \in \Sigma^* \times \mathbb{N}$, $k$ is called the \emph{parameter}.
A parameterized problem is \emph{fixed-parameter tractable} \FPT\ if there exists an algorithm $\mathcal{A}$, a computable function $f$, and a constant $c$ such that given an instance $I=(x,k)$,
$\mathcal{A}$ (called an \FPT\ \emph{algorithm}) correctly decides whether $I \in L$ in time bounded by $f(k) \cdot |I|^c$.
A fundamental concept in parameterized complexity is that of \emph{kernelization}; see~\cite{book-kernels} for a recent book on the topic. A kernelization
algorithm, or just \emph{kernel}, for a parameterized problem $\Pi $ takes an
instance~$(x,k)$ of the problem and, in time polynomial in $|x| + k$, outputs
an instance~$(x',k')$ such that $|x'|, k' \leqslant g(k)$ for some
function~$g$, and $(x,k) \in \Pi$ if and only if $(x',k') \in \Pi$. The function~$g$ is called the \emph{size} of the kernel. A kernel is called \emph{polynomial} (resp. \emph{quadratic, linear}) if the function $g(k)$ is a polynomial (resp. quadratic, linear) function in $k$.

In terms of parameterized enumeration, we refer the reader to~\cite{creignou2017enum,golovach2022refined} for a more comprehensive overview than what we give below. A \textit{parameterized enumeration problem} is a triple $(L, \Sol, \kappa)$ where $L \subseteq \Sigma^*$ is a language, $\Sol: \Sigma^+ \mapsto 2^{\Sigma^*}$ is the set of all viable solutions and $\kappa: \Sigma^* \mapsto \mathbb{N}$ is a parameterization. An instance to a parameterized enumeration problem is a pair $(x, k)$ where $k = \kappa(x)$ and the goal is to produce $\Sol(x)$.
We say that an algorithm $\mathcal{A}$ that takes $(x,k)$ as input is a \TFPT\ algorithm if it outputs $\Sol(x)$ in \FPT\ time.
Naturally, several problems won't have $\Sol(x)$ by of \FPT\ size.
In this case, the best we can hope for is that the \textit{delay} to outputting a new is \FPT. If not only this is the case but also: (i) the time to the first solution, and (ii) the time from the final solution to the halting of the algorithm are also in \FPT, then we say that the algorithm is a \DFPT\ algorithm.
Very recently, Golovach et al.~\cite{golovach2022refined} gave kernelization analogues to \TFPT\ and \DFPT, which they called \textit{fully-polynomial enumeration kernel} (FPE) and \textit{polynomial-delay enumeration kernel} (PDE), respectively.
Formally, an FPE kernel is a pair of algorithms $\mathcal{A}, \mathcal{A}'$ called the \textit{compressor}\footnote{This was named the \textit{kernelization} algorithm in~\cite{golovach2022refined}, but we reserve this term to the pair $\mathcal{A}, \mathcal{A}'$ itself.} and \textit{lifting} algorithms, respectively, where:

\begin{itemize}
    \item Given $(x, k)$, $\mathcal{A}$ outputs $(x', k')$ with $|x'|,k' \leq g(k)$ in time $\poly(|x| + k)$ for some computable $g$.
    \item For each $s \in \Sol(x')$, $\mathcal{A}'$ computes a set $S_s$ in time $\poly(|x| + |x'| + k + k')$ such that $\{S_s \mid s \in \Sol(x')\}$ is a partition of $\Sol(x)$.
\end{itemize}

For PDE kernels, we replace the polynomial (total) time  condition of $\mathcal{A}'$ with \textit{polynomial delay} on $|x| + |x'| + k + k'$.

%% file: np_cubic.tex
\section{(Sub)Cubic graphs}

A result of Chvátal~\cite{chvatal1984recognizing} from the 1980s shows that \pname{Matching Cut} is polynomial-time solvable for subcubic graphs.
Later, Moshi~\cite{matching_cut_moshi} showed that every connected subcubic graph on at least eight vertices has a matching cut.
When dealing with  \pname{Matching Multicut}, the situation is not as simple.
We first show that, if the number of components $\ell$ is part of the input, then \pname{Matching Multicut} is $\NPH$.
However, we are able to prove a Moshi-like result, and show that, if $\ell$ is a parameter, then the problem admits a quasi-linear kernel.

\subsection{NP-hardness}
First, let us show a lemma and some helpful definitions for our construction.

\begin{definition}
	A graph $G$ is indivisible if and only if $G$ has no matching cut. Similarly, a set of vertices $X \subseteq V(G)$ is said to be indivisible if the subgraph of $G$ induced by $X$ is indivisible.
\end{definition}

We remark that the above definition is a conservative notion of togetherness; i.e. we do not require that $X$ is together in every matching cut of $G$, we require it to be together \textit{regardless} of the remainder of the graph that contains it.

\begin{definition}
	Let $X \subset V(G)$ induce a connected subgraph of $G$ with exactly one $u \in X$ such that $N(u) \setminus X \ne \emptyset$. If $|N(u) \setminus X| = 1$ we say $G[X]$ is a pendant subgraph of $G$ and that $X$ induces a pendant subgraph of $G$.
\end{definition}

\begin{lemma}
	\label{lem8}
	Let $I = \{H_1, \ldots, H_k\}$ be a set of maximal indivisible pendant subgraphs of $G$. Let $v_1, \ldots, v_k$ be pairwise distinct vertices so that $N(H_i) = \{v_i\}$.
	If $(G, \ell)$ is a yes-instance for some $\ell > k$, then there exists a  matching multicut $\mathcal{P} = \{P_1, \ldots, P_\ell\}$ with $P_i = V(H_i)$ for $1 \leq i \leq k$. 
\end{lemma}

\begin{proof}
	We prove our claim by induction on $k$. For simplicity, we assume w.l.o.g that all parts of our matching multicuts induce connected graphs of $G$.
	
	For $k=1$, take any matching multicut $\mathcal{P}'=\{P'_1, \ldots, P'_\ell\}$.
	Suppose that $H_1 \subseteq P'_1$. If $H_1 = P'_1$ we are done.
	Otherwise, If $H_1 \subset P'_1$ and $v_1$ does not have a neighbor outside of $P'_1$, then $\mathcal{P} = \{H_1, (P'_1 \setminus H_1), \cup P'_2, P'_3, \ldots, P'_\ell\}$ is a matching multicut with the desired number of parts.
	If $v_1$ has a neighbor $w$ in, say $P'_2$, we can define $\mathcal{P} = \{H_1, P'_1 \cup P'_2 \setminus H_1, \dots, P'_\ell \}$, i.e. we merge components $P'_1$ and $P'_2$, except for $H_1$, which now has a matching edge to $v$ since $H_1$ is a pendant subgraph of $G$.
	
	For the general case, suppose that the lemma holds for all $a < k$.
	By the induction hypothesis, $G$ admits a matching multicut $\mathcal{P}'=\{H_1, \ldots, H_{k-1}, P'_k, \ldots, P'_\ell\}$. Suppose that $H_k \subset P'_k$. If $H_k = P'_k$ we are done.
	Otherwise, since $v_i \neq v_j$ for all $i,j \in [k]$, it follows that $v_k \in P'_k$ and $N(v_k) \cap V(H_i) = \emptyset$. As such, by proceeding as in the base case we can isolate $H_k$ in its own part without gobbling up another $H_i$.
\end{proof}

{\noindent\bf Construction.} To construct our instance $(H, \ell)$  of \pname{Matching Multicut}, we will reduce from and instance $(G, k)$ of \pname{Independent Set} on cubic graphs, which is a well known \NPc\ problem \cite{garey}. 

\begin{figure}[!htb]
\begin{center}
\begin{tikzpicture}[scale=0.7]
	\tikzset{VertexStyle/.append style = {inner sep = \inners, outer sep = \outers,  fill=black, minimum size=2pt}}
    
	\SetVertexLabelOut
	\begin{scope}
		\Vertex[x=2.25, y=1.75, Lpos=180, Math, L={f'_i}]{v_{8}}
		\Vertex[x=5.75, y=3.75, Lpos=0, Math, L={f_i}]{v_{11}}
\SetVertexNoLabel
\Vertex[x=1.5, y=-1.925, Lpos=90, Math, L={v_{0}}]{v_{0}}
\Vertex[x=3.5, y=-1.925, Lpos=90, Math, L={v_{1}}]{v_{1}}
\Vertex[x=4.575, y=-1.925, Lpos=90, Math, L={v_{3}}]{v_{3}}
\Vertex[x=6.575, y=-1.925, Lpos=90, Math, L={v_{4}}]{v_{4}}
\Vertex[x=4, y=0.75, Lpos=90, Math, L={v_{6}}]{v_{6}}
\Vertex[x=4, y=1.75, Lpos=90, Math, L={v_{7}}]{v_{7}}
\SetVertexLabel
\Vertex[x=4, y=2.75, Lpos=180, Math, L={v_{9}}, L={n_i}]{v_{9}}
\Vertex[x=2.5, y=-0.425, Lpos=180, Math, L={v_{2}}, L={B_{u}}]{v_{2}}
\Vertex[x=5.5, y=-0.425, Lpos=0, Math, L={v_{5}}, L={B_{v}}]{v_{5}}
\Vertex[x=4, y=3.75, Lpos=180, Math, L={v_{10}}, L={p_i}]{v_{10}}
\Edge[](v_{0})(v_{1})
\Edge[](v_{2})(v_{0})
\Edge[](v_{1})(v_{2})
\Edge[](v_{3})(v_{4})
\Edge[](v_{4})(v_{5})
\Edge[](v_{5})(v_{3})
\Edge[](v_{2})(v_{6})
\Edge[](v_{6})(v_{5})
\Edge[](v_{10})(v_{9})
\Edge[](v_{9})(v_{7})
\Edge[](v_{7})(v_{6})
\Edge[color=red, style={double}](v_{11})(v_{10})
\Edge[color=red,style={double}](v_{8})(v_{7})
	\end{scope}
\end{tikzpicture}
   \caption{Edge gadget for edge $e_i = uv$. Thick edges are assumed to be in any solution to \pname{Matching Multicut}. \label{fig:reduction}}
\end{center}
\end{figure}
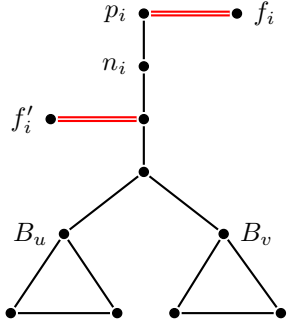

First, for each $u \in V(G)$, create a $K_3$ in $H$, label it as $B_u$, and let $B = \bigcup_{u \in V(G)} B_u$. Suppose that $E(G)$ has been arbitrarily ordered as $\{e_1, \ldots, e_m\}$.
For each edge in order, we add the gadget in Figure~\ref{fig:reduction}. The vertices in $B_u$ and $B_v$ are connected in such a way no vertex has more than three neighbors.

\tikzstyle{red edge}=[-, draw=red, thick]
\tikzstyle{black edge}=[-, thick]

To connect our edge edges, we add an edge between $n_i$ and $p_{i+1}$ for every $i \in [m-1]$ and between $n_m$ and $p_1$.
With this the subgraph of $H$ induced by the $n_i$'s and $p_i$'is a cycle on $2m$ vertices. Finally, we set $\ell = 2m + k+1$.

\begin{theorem} \label{thm:np-hard}
    $(H, 2|E(G)| + k + 1)$ is a \YES-instance of \pname{Matching Multicut} if and only if $G$ has an independent set of size at least $k$. Moreover, \pname{Matching Multicut} is \NPc \, in subcubic graphs.
\end{theorem}

\begin{proof}
Let $X = \{x_1, \ldots, x_k\}$ be an independent set in $G$. The partition $\mathcal{P} = \{f_1, f'_1, \ldots, f_m, f'_m, B_{x_1}, \ldots, B_{x_k}, R\}$, where $R$ are all other vertices of $H$ not explicitely listed in $\mathcal{P}$.
Since $X$ is an independent set of $G$, it holds that, for each edge gadget, at most one of $B_u, B_v$ was added to $\mathcal{P}$, so it immediately follows that no vertex has two neighbors outside of its own part. Moreover, for every $B_u$, it is either isolated as a single component in $\mathcal{P}$ or it is part of $R$. Consequently, $\mathcal{P}$ is a matching multicut of size $2|E(G)| + k +1 1$.
	
For the other direction, let $I$ be the set of all pendant vertices and $\mathcal{P}$ be a matching multicut of $H$; by Lemma~\ref{lem8}, we may assume w.l.o.g. that every pendant vertex has its only edge colored as part of the cut, i.e. each pendant vertex is in its own part of $\mathcal{P}$.
Additionally, the pendant edges across all gadgets form a maximal matching in $H \setminus (B \cup I)$.
Therefore, we have $R = H \setminus (B \cup I)$ contained in a single set of $\mathcal{P}$.
Hence, the remaining $k$ sets are some of the triangles of $B$. Note that if $uv \in E(G)$, either $B_u$ or $B_v$ is an element of $\mathcal{P}$, but never both. Therefore, $X = \{u \mid B_u \in \mathcal{P}\}$ is an independent set of $G$ of size at least $k$.

The final claim follows immediately from the fact that \pname{Independent Set} on cubic graphs is \NPH\ and that verifying if a partition is a matching multicut is naively done in polynomial time.
\end{proof}

\begin{figure}[!htb]
 \tikzset{hide labels/.style={every label/.append style={text opacity=0}}}

\begin{center}
    \centering
	\begin{tikzpicture}[hide labels, scale=0.7]
	\GraphInit[unit=3,vstyle=Normal]
	\SetVertexNormal[Shape=circle, FillColor=black, MinSize=2pt]
	\tikzset{VertexStyle/.append style = {inner sep = \inners, outer sep = \outers}}
    
	\SetVertexLabelOut
	
	\begin{scope}
	\Vertex[x=-5,y=3,Lpos=90, Math, L={v_1}]{v_1}
	\Vertex[x=-2, y=3,Lpos=90, Math, L={v_2}]{v_2}
	\Vertex[x=-2, y=0,Lpos=90, Math, L={v_3}]{v_3}
	\Vertex[x=-5, y=0,Lpos=0, Math, L={v_4}]{v_4}
	\Vertex[x=-3.5, y=1.5,Lpos=0, Math, L={v_5}]{v_5}
	\Edge(v_1)(v_2)
	\Edge(v_1)(v_4)
	\Edge(v_3)(v_4)
	\Edge(v_2)(v_3)
        \Edge(v_5)(v_3)
        \Edge(v_5)(v_4)
        \Edge(v_5)(v_1)
	\end{scope}
\end{tikzpicture}
\caption{Indivisible pendant subgraph. \label{fig:ind_pend}}
\end{center}
\end{figure}
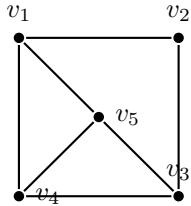

We can show in a very similar manner that \pname{Matching Multicut} is \NPH\ for cubic graphs. To do this, we replace the pendant vertices of $H$ with the indivisible graph in Figure~\ref{fig:ind_pend}.
The remainder of the argument follows as in the proof of the previous theorem.

\subsection{Quasi-linear kernel}

We now present a quasi-linear kernel for the \pname{Matching Multicut} problem in the case where the number of partitions $\ell$ is a parameter. In order to construct the kernel, we extend Moshi's result~\cite{matching_cut_moshi} and show the following theorem.

\begin{theorem}\label{thm:quasi_lin_kern}
    Let $G$ be a connected graph with $\Delta(G) \leq 3$. If $|V(G)| = \Omega(\ell \log^2 \ell)$, then $G$ has a matching multicut that partitions the graph into $\ell$ parts.
\end{theorem}

The key for proving Theorem~\ref{thm:quasi_lin_kern} is to find a sufficiently large collection of vertex-disjoint cycles and construct a partition using some of them. In order to find these cycles, we first need to deal with vertices of degree at most 2.

\begin{lemma}\label{lem:deg1}
    Let $G$ be a connected subcubic graph, and let $V_1$ denote the vertices of $G$ with degree 1. If $|V_1| \geq 3\ell$, then $G$ has a matching multicut that partitions the graph into $\ell$ parts.
\end{lemma}
\begin{proof}
    Construct a matching $M$ by greedily choosing edges that contain a vertex from $V_1$. Because $\Delta(G) \leq 3$, $|M| \geq \frac{1}{3} |V_1| \geq \ell$. Moreover, $G-M$ contains at least $|M|$ components with an isolated vertex.
\end{proof}

\begin{lemma}\label{lem:deg2}
    Let $G$ be a connected subcubic graph with $|V(G)| = \Omega(\ell)$, and let $V_2$ denote the vertices of $G$ with degree 2. If $|V_2| \geq \frac{9}{10} |V(G)|$, then $G$ has a matching multicut that partitions the graph into $\ell$ parts.
\end{lemma}

\begin{proof}
We call a set of 4 distinct vertices $\{u', u, v, v'\}$ a subdivided edge if $d(u) = d(v) = 2$, and $\{u'u, uv, vv'\} \subseteq E(G)$. The idea is that because we have a large fraction of degree 2 vertices, there are many subdivided edges in $G$.

More formally, let $V_2$ be the set of vertices of degree $2$, let $n = |V(G)|$, and let $f = \frac{|V_2|}{n} \geq \frac{9}{10}$. There are at most $3|V(G) \setminus V_2| = 3(1-f)n$ edges crossing from $V_2$ to its complement. Therefore, there are at least $\frac{1}{2}(2|V_2| - 3|V(G) \setminus V_2|) = \frac{1}{2}(5f-3)n$ edges inside the set $V_2$, which is at least $\frac{3}{4}n$ if $f \geq \frac{9}{10}$.

We want to use the edges inside $V_2$ to construct subdivided edges of $G$. However, it can be the case that for an edge $uv$ in $G[V_2]$, there exists a vertex $w \in V(G) \setminus V_2$ forming a triangle with $u$ and $v$, thus not forming a subdivided edge. For this reason, we look for paths with 3 vertices inside $V_2$. Notice that because we have more than $\frac{1}{2}|V_2|$ edges inside $V_2$, $G[V_2]$ must contain at least one $P_3$. Actually, it is easy to see that at least $2(\frac{3}{4} - \frac{1}{2})|V_2|$ vertices have degree 2 inside $V_2$ (by degree counting), so we have at least $|V_2|/2$ paths with 3 vertices, implying at least $n/3$ subdivided edges exist in $G$.

We greedily construct the matching multicut $M$ by taking subdivided edges and marking vertices that cannot be taken. More formally, let $M = \emptyset$ be the edges of the current matching multicut and $S = \emptyset$ be the set of marked vertices. While there exists a subdivided edge $(u', u, v, v')$ with $u', u, v, v' \notin S$, add edges $u'u$ and $vv'$ to $M$ and mark the subdivided edge, i.e., $M := M \cup \{u'u, vv'\}$ and $S := S \cup \{u', u, v, v'\}$.

At each step, we add to $M$ edges with non-marked endpoints, thus $M$ indeed forms a matching. Moreover, in each step, we eliminate at most $7$ subdivided edges (the worst case occurs when the subdivided edge contains the 4 central vertices of a $P_{10}$). By the end, $|M|$ is at least $\frac{1}{7}$ of the total number of subdivided edges. Therefore, we require $|V(G)| \geq 21 \ell$.
\end{proof}

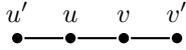
\begin{figure}[!htb]
 \tikzset{hide labels/.style={every label/.append style={text opacity=0}}}

\begin{center}
    \centering
	\begin{tikzpicture}[scale=0.7]
	\GraphInit[unit=3,vstyle=Normal]
	\SetVertexNormal[Shape=circle, FillColor=black, MinSize=2pt]
	\tikzset{VertexStyle/.append style = {inner sep = \inners, outer sep = \outers,  fill=black, minimum size=2pt}}
    
	\SetVertexLabelOut
	
	\begin{scope}

	\Vertex[x=1,y=0,Lpos=90, Math, L={u'}]{v_1}
	\Vertex[x=2, y=0,Lpos=90, Math, L={u}]{v_2}
	\Vertex[x=3, y=0,Lpos=90, Math, L={v}]{v_3}
	\Vertex[x=4, y=0,Lpos=90, Math, L={v'}]{v_4}
	\Edge(v_1)(v_2)
	\Edge(v_2)(v_3)
	\Edge(v_3)(v_4)
	\end{scope}
\end{tikzpicture}
\caption{Subdivided edge. \label{fig:sub_edge}}
\end{center}
\end{figure}

Assume that $G$ is a subcubic graph that does not satisfy the degree conditions of Lemmas \ref{lem:deg1} and \ref{lem:deg2}. That is, $G$ does not have many degree-1 vertices nor a large proportion of degree-2 vertices. Our strategy to find a matching multicut for $G$ involves using disjoint cycles. The intuition here is that if a cycle $C$ is entirely contained within a part of a partition of $G$, each vertex $v \in C$ will have at most one edge crossing the partition. Therefore, cycles are a good starting point for partitioning the matching multicut. However, first we need to find disjoint cycles. For this purpose, we utilize a theorem due to Simonovits \cite{simonovits1967new}, which is a precise version of the well-known Erdős–Pósa theorem \cite{Erdös_Pósa_1965} in the context of subcubic graphs.

\begin{theorem}[Simonovits '67] \label{thm:simonovits}
Let $G$ be a connected graph with $\delta(G) \geq 2$. Let $V_{\geq 3}$ be the set of vertices of $G$ with degree at least 3. Then, $G$ has at least $|V_{\geq 3}| / (4\log |V_{\geq 3}|)$ vertex disjoint cycles.
\end{theorem}

It is worth mentioning that there exists an algorithmic approximation of Theorem \ref{thm:simonovits} due to Brandstädt and Voss~\cite{BRANDSTADT1996197}. Therefore, all the theorems presented in this subsection are constructive and can be used to find a matching multicut of subcubic graphs.

Before moving to the main theorem of this subsection, we make some observations about the neighborhood of subcubic graphs. Let $N(v)$ be the set of vertices of $G$ adjacent to $v$, and let $N[v] = N(v) \cup \{v\}$. More generally, let $N(S)$ be the set of vertices of $G - S$ that are adjacent to some vertex in $S$, and let $N[S] = N(S) \cup S$. We call $N(S)$ the open neighborhood and $N[S]$ the closed neighborhood. We denote by $N^2[S] := N[N[S]]$ the closed square neighborhood. Notice that for subcubic graphs, $|N^2[S]| \leq 10|S|$.

\begin{proof}[Proof of Theorem \ref{thm:quasi_lin_kern}]
    Let $G$ be a graph satisfying the conditions of Theorem $\ref{thm:quasi_lin_kern}$. Let $V_1(G)$, $V_2(G)$, and $V_3(G)$ be the subsets of vertices of $G$ with degrees 1, 2, and 3, respectively. If $G$ satisfies the conditions of Lemmas \ref{lem:deg1} or \ref{lem:deg2}, we are done. Now we can safely assume that $|V_1(G)| < 3\ell$ and $|V_2(G)| < \frac{9}{10} |V(G)|$.
    
    Let $G'$ be the graph obtained from $G$ after recursively removing degree 1 vertices. Notice that each time a degree 1 vertex is removed, a vertex moves from $V_2$ to $V_1$ or from $V_3$ to $V_2$. In both cases, the difference $|V_3| - |V_1|$ remains invariant; therefore, $|V_3(G')| \geq |V_3(G)| - |V_1(G)|$. Notice that any matching multicut of $G'$ is also a matching multicut of $G$. Assume again that $G'$ does not satisfy Lemma \ref{lem:deg2}; in particular, this implies that $|V_3(G)| \geq |V(G)|/10$.

    Now, $G'$ satisfies the conditions of Theorem \ref{thm:simonovits}. Let $\{C_1, \dots, C_k\}$ be a collection of $k = |V_3(G')| / (4 \log |V_3(G')|) = \Omega(\ell \log \ell)$ vertex-disjoint sets such that $G[C_i]$ is a cycle. By giving a lower bound for the value of $k$, we also give a lower bound for $|V(G)|$. Later in the proof, we will need $k$ such that $k^2 \geq c \ell |V(G')|$, but notice that there always exists a constant $c' > c$ such that if $|V(G)| \geq c' \ell \log^2 \ell$, the lower bound on $k^2$ is satisfied.
    
    For each set of vertices $C_i$, if there is $v \in V(G') \setminus C_i$ with $|N(v) \cap C_i| \geq 2$, add $v$ to $C_i$, that is, $C_i := C_i \cup \{v\}$. Notice that with this process, every vertex inside $C_i$ has at least two neighbors inside $C_i$, therefore, $E(C_i, V(G') \setminus C_i)$ forms a matching cut.
 
    We construct the matching multicut greedily. Let $M := \emptyset$ be the initial matching multicut and let $S := \emptyset$ be a collection of marked vertices. Assume that the sets $C_i$ are ordered by size with $|C_1| \leq \dots \leq |C_k|$. Let $C_i$ be a set in the first half of this ordering with no vertex marked, i.e., $C_i \cap S = \emptyset$. Add the edges with exactly one endpoint in $C_i$ to $M$ and mark $N^2[C_i]$, that is, $M := M \cup E(C_i, V(G') \setminus C_i)$ and $S := S \cup N^2[C_i]$. If no such $C_i$ exists in the first half of the ordering, stop the process. We claim that in the end, $M$ is indeed a matching multicut.

\begin{lemma}
    If $M$ is a set of edges constructed as above, then $M$ is a matching multicut that divides $G'$ into at least $\ell$ parts.
\end{lemma}

   It is easy to see that $M$ is indeed a matching. Assuming otherwise, then there is a vertex $v$ with two edges from $M$ containing $v$. By previous observations, $v$ must not belong to any set $C_i$ whose border was added to $M$, thus $v \in V(G') \setminus (C_1 \cup \dots \cup C_k)$. If $|N(v) \cap C_i| \geq 2$, $v$ would already have been added to $C_i$, so this cannot be the case. Hence, there are distinct sets $C_i$ and $C_j$ chosen in the algorithm with $|N(v) \cap C_i|, |N(v) \cap C_j| \geq 1$. Assume that $C_i$ was chosen before $C_j$. As $v$ is adjacent to a vertex of $C_i$ and a vertex of $C_j$, there is a vertex of $C_j$ in $N^2[C_i]$, which means that this vertex should have been marked, implying that this situation also cannot happen. We conclude that there is no vertex $v$ with two edges from $M$ containing $v$.

    Now, we just need to check that during the process at least $\ell$ sets $C_i$ were chosen so that the edges $E(C_i, V(G') \setminus C_i)$ were added to $M$. If this does not occur, we have that the size of the marked vertices $|S|$ is bounded:
\begin{equation*}
    |S| \leq (\ell - 1) \max_{i \leq k/2} \{|N^2[C_i]|\} \leq 10(\ell - 1) |C_{k/2}| \leq 10(\ell - 1) \frac{|V(G)|}{k/2} < \frac{k}{2}
\end{equation*}
In the first inequality, we are assuming the worst case where we have always added the largest squared neighbourhood. The second inequality follows from our previous bound on the size of this squared closed neighbourhood. The third inequality holds because the average size of the $k/2$ largest sets $C_i$ is $2|V(G)|/k$, and the set $C_{k/2}$ has a size below this average. The last inequality follows from our lower bound on $k$.

This concludes the proof by showing that $M$ indeed divides $G'$ into at least $\ell$ components, so it is a matching multicut.
\end{proof}

Notice that in the proof, we choose $|V(G)|$ in order to establish a lower bound on $k^2$. We do not explicitly specify the choice of the constant $c'$ such that $|V(G)| \geq c' \ell \log^2 \ell$. However, through a simple computation, it can be shown that $c' = 10^6$ is sufficient. We have not attempted to minimize the constants, but we believe that the value of $c'$ can be significantly reduced.

It follows from Theorem \ref{thm:quasi_lin_kern} that if we want to ask for a matching multicut that divides an $n$-vertex subcubic graph into $\ell = \mathcal{O}(n/\log^2 n)$ parts, the answer is trivially yes. On the other hand, Theorem \ref{thm:np-hard} provides a construction of a subcubic graph and shows that it is NP-hard to determine if this graph has a matching multicut that divides it into $\Theta(n)$ parts. We leave it as an open question if it is possible to improve the asymptotic bound given by Theorem \ref{thm:quasi_lin_kern}.

%% file: exact.tex
\section{Exact Exponential Algorithm}

We now turn our attention to developing an exact exponential algorithm through a similar approach used in \cite{matching_cut_ipec}.
For more on this type of algorithm and its associated terminology, we refer the reader to~\cite{exact_exponential_algorithms}.
Our algorithm consists of four stopping rules, seven reduction and nine branching rules.
At every step of the algorithm we have the sets $\{A_1, \dots, A_\ell, F\}$ such that $\varphi = \{A_1, \dots, A_\ell\}$ (unless any stopping rule is applicable) is a matching $\ell$-multicut of the vertices of $V(G) \setminus F$.
For simplicity, we assume that $\delta(G) \geq 2$.
Most of the arguments presented here work with slight modifications to graphs of minimum degree one, but they would unnecessarily complicate the description of the algorithm.

\begin{itemize}
	\item[S1] If there is some $v \in F$ and $i,j \in [\ell]$ such that $\deg_{A_i}(v) \geq 2$ and $\deg_{A_j}(v) \geq 2$, STOP: there is no matching $\ell$-multicut extending $\varphi$.
	\item[S2] If there is a vertex $v \in F$ with neighbors in three different parts of $\varphi$, STOP: there is no matching $\ell$-multicut extending $\varphi$.
	\item[S3] If there is an edge $uv$ with $u \in A_i$ and $v \in A_j$ such that $N(u) \cap N(v) \cap F \neq \emptyset$, STOP: there is no matching $\ell$-multicut extending $\varphi$.
	\item[S4] If there is some $v \in A_i$ with two neighbors outside of $A_i \cup F$, STOP: there is no matching $\ell$-multicut extending $\varphi$.
\end{itemize}

Intuitively, stopping rules are applicable whenever a bad decision has been made by the branching algorithm and we must prune that branch.
The following are our reduction rules, and are useful for cleaning up an instance after a branching step has been performed.

\begin{itemize}
	\item[R1] If there exists some $v \in A_i$ such that $N(v) \supseteq \{x,y\}$ and $x,y \in F$ and $xy \in E(G)$, add $x,y$ to $A_i$.
	\item[R2] If there exists $v \in F$ and a unique $i \in [\ell]$ with $\deg_{A_i}(v) \geq 2$, add $v$ to $A_i$.
	\item[R3] For every edge $uv$ with $u \in A_i$ and $v \in A_j$, add $N(u) \cap F$ to $A_i$ and $N(v) \cap F$ to $A_j$.
	\item[R4] If there is a pair $u,v \in F$ with $N(u) = N(v) = \{x, y\}$ with $x \in A_i$ and $y \in A_j$, add $u$ to $A_i$ and $v$ to $A_j$.
	\item[R5] If there is a pair $u,v \in F$ with $N(u) = N(v) = \{x, y\}$ with $x \in A_i$ and $y \in F$, add $u$ to $A_i$.
	\item[R6] If there is a vertex $v \in F$ with $N(v) = \{x,y\}$, $x \in A_i$, $y \in A_j$, $N(x) \subseteq A_i \cup \{v\}$, and $N(y) \subseteq A_j \cup \{v\}$, add $u$ to $A_i$.
	\item[R7] If there are vertices $u,v,w \in F$ with $\deg(u) = \deg(v) = \deg(w)$ arranged as in Figure~\ref{fig:r7}, add $\{u,v\}$ to $A_i$ and $w$ to $A_j$.
	
	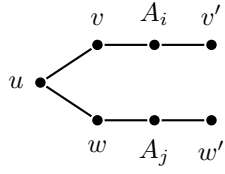
\begin{figure}[!htb]
		\centering
		\begin{tikzpicture}[scale=0.5]
		\GraphInit[unit=3,vstyle=Normal]
		\SetVertexNormal[Shape=circle, FillColor=black, MinSize=2pt]
		\tikzset{VertexStyle/.append style = {inner sep = \inners, outer sep = \outers}}
		\Vertex[x=0, y=0, LabelOut, Lpos=180, Math]{u}
		\SetVertexNoLabel
		\foreach \y in {-1, 1} {
			\foreach \x in {0,1,2} {
				\pgfmathsetmacro{\xp}{1.5+1.5*\x}
				\Vertex[x=\xp, y=\y, LabelOut, Lpos=90]{v\y\x}
			}
			\foreach \x in {0,1} {
				\pgfmathtruncatemacro{\xp}{1+\x}
				\Edge(v\y\x)(v\y\xp)
			}
			\Edge(u)(v\y0)
		}
		\tikzset{AssignStyle/.append style = {above=4pt}}
		\AssignVertexLabel{v1}{$v$,$A_i$,$v'$}
		\tikzset{AssignStyle/.append style = {below=8pt}}
		\AssignVertexLabel{v-1}{$w$,$A_j$,$w'$}
		\end{tikzpicture}
		\caption{\label{fig:r7} Configuration for rule R7.}
	\end{figure}
\end{itemize}

\begin{figure}[!htb]
	\centering
	\begin{tikzpicture}[scale=0.5]
	\GraphInit[unit=3,vstyle=Normal]
	\SetVertexNormal[Shape=circle, FillColor=black, MinSize=2pt]
	\tikzset{VertexStyle/.append style = {inner sep = \inners, outer sep = \outers}}
	\SetVertexLabelOut
	
	\newcommand{\brule}{1}
	\begin{scope}[shift={(-10,5)}]
	\Vertex[x=0,y=0,Lpos=90, Math, L={v_1}]{v_1\brule}
	\Vertex[x=-1.5, y=0,Lpos=90, Math, L={A_i}]{a_i\brule}
	\Vertex[x=-3, y=0,Lpos=90, Math, L={v_2}]{v_2\brule}
	\Vertex[x=1.5, y=1,Lpos=0, Math, L={v_3}]{v_3\brule}
	\Vertex[x=1.5, y=-1,Lpos=0, Math, L={v_4}]{v_4\brule}
	\Edge(v_1\brule)(a_i\brule)
	\Edge(v_1\brule)(v_3\brule)
	\Edge(v_1\brule)(v_4\brule)
	\Edge(a_i\brule)(v_2\brule)
	\node at (-0.5,-1) {(B\brule)};
	\end{scope}
	
	\renewcommand{\brule}{2}
	\begin{scope}[shift={(0,5)}]
	\Vertex[x=0,y=0,Lpos=90, Math, L={v_1}]{v_1\brule}
	\Vertex[x=-1.5, y=0,Lpos=90, Math, L={A_i}]{a_i\brule}
	\Vertex[x=-3, y=0,Lpos=90, Math, L={v_2}]{v_2\brule}
	\Vertex[x=1.5, y=1,Lpos=0, Math, L={A_j}]{a_j\brule}
	\Vertex[x=1.5, y=-1,Lpos=0, Math, L={v_4}]{v_4\brule}
	\Edge(v_1\brule)(a_i\brule)
	\Edge(v_1\brule)(a_j\brule)
	\Edge(v_1\brule)(v_4\brule)
	\Edge(a_i\brule)(v_2\brule)
	\node at (-0.5,-1) {(B\brule)};
	\end{scope}
	
	\renewcommand{\brule}{3}
	\begin{scope}[shift={(8,5)}]
	\Vertex[x=0,y=0,Lpos=90, Math, L={v_1}]{v_1\brule}
	\Vertex[x=-1.5, y=0,Lpos=90, Math, L={v_2}]{v_2\brule}
	\Vertex[x=1.5, y=1,Lpos=0, Math, L={A_i}]{a_i\brule}
	\Vertex[x=1.5, y=-1,Lpos=0, Math, L={A_j}]{a_j\brule}
	\Edge(v_1\brule)(a_i\brule)
	\Edge(v_1\brule)(a_j\brule)
	\Edge(v_1\brule)(v_2\brule)
	\node at (0,-1) {(B\brule)};
	\end{scope}
	
	\renewcommand{\brule}{4}
	\begin{scope}[shift={(-8.5,0)}]
	\Vertex[x=0,y=0,Lpos=90, Math, L={v_1}]{v_1\brule}
	\Vertex[x=1.5,y=0,Lpos=0, Math, L={v_4}]{v_4\brule}
	\Vertex[x=-1.5, y=1, Lpos=90, Math, L={v_2}]{v_2\brule}
	\Vertex[x=-3, y=1, Lpos=90, Math, L={A_i}]{a_i\brule}
	\Vertex[x=-4.5, y=1, Lpos=90, Math, L={v_2'}]{v_2p\brule}
	\Vertex[x=-1.5, y=-1, Lpos=270, Math, L={v_3}]{v_3\brule}
	\Vertex[x=-3, y=-1, Lpos=270, Math, L={A_j}]{a_j\brule}
	\Vertex[x=-4.5, y=-1, Lpos=270, Math, L={v_3'}]{v_3p\brule}
	
	\node at (0.5,-1) {(B\brule)};
	\Edge(v_1\brule)(v_4\brule) \Edges(v_2p\brule,a_i\brule,v_2\brule,v_1\brule,v_3\brule,a_j\brule,v_3p\brule)
	\end{scope}
	
	\renewcommand{\brule}{5}
	\begin{scope}[shift={(0,0)}]
	\Vertex[x=0,y=0,Lpos=90, Math, L={v_1}]{v_1\brule}
	\Vertex[x=1.5,y=0,Lpos=0, Math, L={A_j}]{a_j\brule}
	\Vertex[x=-1.5, y=1, Lpos=90, Math, L={v_2}]{v_2\brule}
	\Vertex[x=-3, y=1, Lpos=90, Math, L={A_i}]{a_i\brule}
	\Vertex[x=-4.5, y=1, Lpos=90, Math, L={v_2'}]{v_2p\brule}
	\Vertex[x=-1.5, y=-1, Lpos=270, Math, L={v_3}]{v_3\brule}
	\Vertex[x=-3, y=-1, Lpos=270, Math, L={A_i}]{a_ip\brule}
	\Vertex[x=-4.5, y=-1, Lpos=270, Math, L={v_3'}]{v_3p\brule}
	
	\node at (0.5,-1) {(B\brule)};
	\Edge(v_1\brule)(a_j\brule) \Edges(v_2p\brule,a_i\brule,v_2\brule,v_1\brule,v_3\brule,a_ip\brule,v_3p\brule)
	\end{scope}
	
	\renewcommand{\brule}{6}
	\begin{scope}[shift={(8.5,0)}]
	\Vertex[x=0,y=0,Lpos=90, Math, L={v_1}]{v_1\brule}
	
	\Vertex[x=-1.5, y=1, Lpos=90, Math, L={v_2}]{v_2\brule}
	\Vertex[x=-3, y=1, Lpos=90, Math, L={A_i}]{a_i\brule}
	\Vertex[x=-4.5, y=1, Lpos=90, Math, L={v_2'}]{v_2p\brule}
	\Vertex[x=-1.5, y=-1, Lpos=270, Math, L={v_3}]{v_3\brule}
	\Vertex[x=-3, y=-1, Lpos=270, Math, L={A_i}]{a_ip\brule}
	\Vertex[x=-4.5, y=-1, Lpos=270, Math, L={v_3'}]{v_3p\brule}

	\Vertex[x=1.5, y=1, Lpos=90, Math, L={v_4}]{v_4\brule}
	\Vertex[x=3, y=1, Lpos=90, Math, L={A_j}]{a_j\brule}
	\Vertex[x=4.5, y=1, Lpos=90, Math, L={v_4'}]{v_4p\brule}
	\Vertex[x=1.5, y=-1, Lpos=270, Math, L={v_5}]{v_5\brule}
	\Vertex[x=3, y=-1, Lpos=270, Math, L={A_j}]{a_jp\brule}
	\Vertex[x=4.5, y=-1, Lpos=270, Math, L={v_5'}]{v_5p\brule}
	
	\node at (0,-1) {(B\brule)}; \Edges(v_2p\brule,a_i\brule,v_2\brule,v_1\brule,v_3\brule,a_ip\brule,v_3p\brule) \Edges(v_4p\brule,a_j\brule,v_4\brule,v_1\brule,v_5\brule,a_jp\brule,v_5p\brule)
	\end{scope}
	
	\renewcommand{\brule}{4}
	\begin{scope}[shift={(-8.5,-6)}]
	\Vertex[x=0,y=0,Lpos=90, Math, L={v_1}]{v_1\brule}
	\Vertex[x=1.5,y=0,Lpos=0, Math, L={v_4}]{v_4\brule}
	\Vertex[x=-1.5, y=1, Lpos=90, Math, L={v_2}]{v_2\brule}
	\Vertex[x=-3, y=1, Lpos=90, Math, L={A_i}]{a_i\brule}
	\Vertex[x=-4.5, y=1, Lpos=90, Math, L={v_2'}]{v_2p\brule}
	\Vertex[x=-1.5, y=-1, Lpos=270, Math, L={v_3}]{v_3\brule}
	\Vertex[x=-3, y=-1, Lpos=270, Math, L={A_i}]{a_j\brule}
	\Vertex[x=-4.5, y=-1, Lpos=270, Math, L={v_3'}]{v_3p\brule}
	
	\node at (0.5,-1) {(B\brule')};
	\Edge(v_1\brule)(v_4\brule) \Edges(v_2p\brule,a_i\brule,v_2\brule,v_1\brule,v_3\brule,a_j\brule,v_3p\brule)
	\end{scope}
	
	\renewcommand{\brule}{7}
	\begin{scope}[shift={(-1,-6)}]
	\Vertex[x=0,y=0,Lpos=90, Math, L={v_1}]{v_1\brule}
	\Vertex[x=-1.5, y=0,Lpos=90, Math, L={A_i}]{a_i\brule}
	\Vertex[x=1.5, y=0,Lpos=90, Math, L={A_j}]{a_j\brule}
	\Vertex[x=-2.5, y=-1,Lpos=270, Math, L={v_2}]{v_2\brule}
	\Vertex[x=2.5, y=-1,Lpos=270, Math, L={v_3}]{v_3\brule}
	\node at (0,-1) {(B\brule)};
	\Edge(v_1\brule)(a_i\brule)
	\Edge(v_1\brule)(a_j\brule)
	\Edge(v_2\brule)(a_i\brule)
	\Edge(v_3\brule)(a_j\brule)
	\end{scope}
	
	\renewcommand{\brule}{8}
	\begin{scope}[shift={(8.5,-6)}]
	\Vertex[x=0,y=0,Lpos=90, Math, L={v_1}]{v_1\brule}
	\Vertex[x=-1.5, y=0,Lpos=90, Math, L={A_i}]{a_i\brule}
	\Vertex[x=1.5, y=0,Lpos=90, Math, L={v_3}]{v_3\brule}
	\Vertex[x=-2.5, y=-1,Lpos=270, Math, L={v_2}]{v_2\brule}
	\Vertex[x=2.5, y=-1,Lpos=270, Math, L={A_j}]{a_j\brule}
	\node at (0,-1) {(B\brule)};
	\Edge(v_1\brule)(a_i\brule)
	\Edge(v_1\brule)(v_3\brule)
	\Edge(v_2\brule)(a_i\brule)
	\Edge(a_j\brule)(v_3\brule)
	\end{scope}
	\end{tikzpicture}
	\caption{\label{fig:nested_branches} Branching configurations for \pname{Matching Multicut}.}
\end{figure}
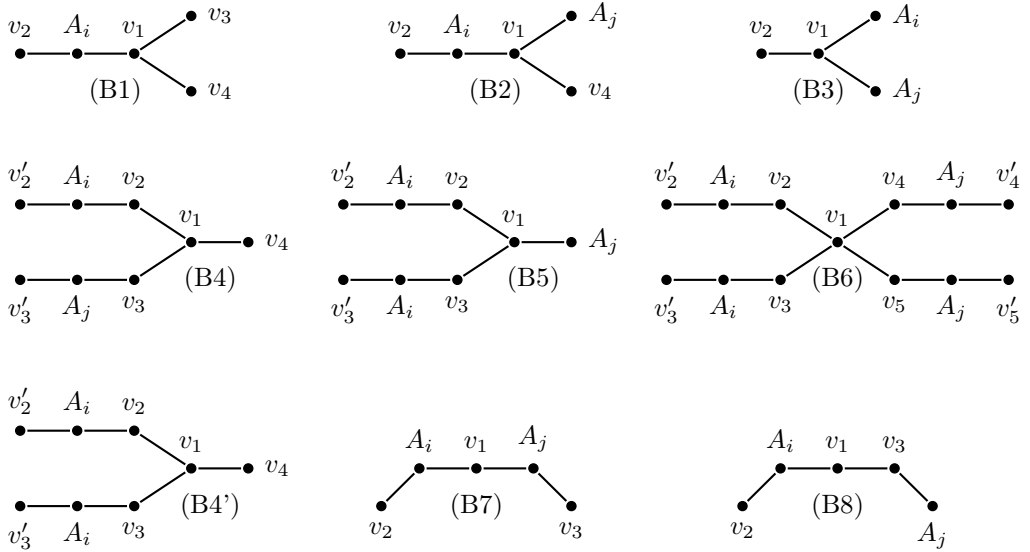

For our branching rules, we follow the configurations given by Figure~\ref{fig:nested_branches}, and always branch on vertex $v_1$.
We set the size of the instance as the size of the set $F$, that is, how many free vertices are not assigned to any part.

\begin{itemize}
	\item[B1] If we put $v_1$ in $A_j$, $j \neq i$, we infer that $v_3$, and $v_4$ must also be added to $A_j$, and that $v_2$ must be added to $A_i$.
	Otherwise, $v_1$ is in $A_i$, which does not given us any additional information.
	Our branching vector is, thus, of the form $\{1\} \times \{4\}^{\ell-1}$.
	
	\item[B2] Note that $v_1$ must be placed in either $A_i$ or $A_j$.
	For the first case, we conclude that $v_4$ must also be in $A_i$, while for the later, $v_4$ must be added to $A_j$ and $v_2$ to $A_i$, yielding the branching vector $(2, 3)$ and the branching factor $1.3247$.
	
	\item[B3] Again, $v_1$ is in either $A_i$ or $A_j$.
	In either case, we conclude that $v_2$ must be in the same part as $v_1$, resulting in the branching vector $(2,2)$, which has a branching factor of $\sqrt{2}$.
	
	\item[B4 and B4'] By adding $v_1$ to $A_i$, we conclude that $v_2$ must also be placed in $A_i$; a similar analysis is performed when $v_1$ is added to $A_j$.
	Otherwise, if we add $v_1$ to $A_k$, at most one of $v_2$ and $v_3$ may be added to a set different from $A_k$.
	If both are in $A_k$, we have that $v_2'$ belongs in $A_i$ and $v_3'$ in $A_j$.
	Otherwise, if $v_2$ is added to $A_i$, we conclude that $v_3,v_4$ belong in $A_k$ and that $v_3'$ belongs in $A_j$; similarly if $v_3$ is assigned to $A_j$.
	This results in a branching vector of the form $\{2\}^2 \times \{5\}^{3\ell-6}$, with unique positive real root of the polynomial associated with it satisfying $\alpha_\ell \leq \sqrt[\leftroot{2} 3]{\ell}$.
	Rule B4' clearly has a better branching factor than B4, but rule B4 dominates the running time of B4'.
	
	\item[B5] In case we assign $v_1$ to $A_i$, we have that both $v_2$, and $v_3$ must also be in $A_i$;
	if $v_1$ is assigned to $A_j$, nothing else can be inferred;
	for all other $A_k$, we have that both $v_2$ and $v_3$ must be assigned to $A_k$, and that $v_2'$ and $v_3'$ belong in $A_i$.
	The branching vector for this rule is given by $\{1\} \times \{3\} \times \{5\}^{\ell-2}$, which, for large values of $\ell$, has a branching factor of at most $\sqrt[\leftroot{2} 3]{\ell}$.
	Again, it can be verified that, for each $\ell$, it holds that the branching factor for this rule is $\leq \sqrt[\leftroot{2} 3]{\ell}$.
	
	\item[B6] If $v_1$ is assigned to $A_i$ (resp. $A_j$), we have that both $v_2$, and $v_3$ (resp. $v_4$, and $v_5$) must also be assigned to $A_i$ (resp. $A_j$);
	otherwise, for every other $A_k$, either $\{v_{p}\}_{p \in [5]}$ belongs to $A_k$, in which case the vertices $\{v'_{p}\}_{p \in \{2,3,4,5\}}$ are assigned to the same set as their neighbor, or at most one $v_p \in \{v_2, v_3, v_4, v_5\}$ is not assigned to $A_k$, in which case the set to which $v'_p$ should be assigned is not determined.
	This rule produces a branching vector of the form $\{3\}^2 \times \{8\}^{4\ell - 8} \times \{9\}^{\ell - 2}$, 
	
	\item[B7] Once again, we only have two options for $v_1$.
	So, if $v_1$ is added to $A_i$, we have that $v_3$ must be added to $A_j$, otherwise $v_1$ is added to $A_j$ and $v_2$ to $A_i$.
	This rule's branching vector is $(2, 2)$, with factor equal to $\sqrt{2}$.
	
	\item[B8] If $v_1$ is assigned to $A_i$, we are done; otherwise, if $v_1$ is assigned to $A_k$, with $k \neq i$, we have that $v_2$ belongs in $A_i$ and $v_3$ in $A_i$.
	This yields the branching vector $\{1\} \times \{3\}^{\ell-1}$, and branching factor $\sqrt[\leftroot{2} 3]{\ell} \leq \alpha_\ell \leq \sqrt{\ell}$.
\end{itemize}

Given all of the above rules, we must show that, if none of them are applicable, we have a matching $\ell$-multicut.
In order to do so, we require some additional definitions: let $A_i' = \{v \in A_i \mid \deg_F(v) \geq 2\}$,  $F_i' = F \cap N(A_i')$, $F_i'' = \{v \in F \mid \deg_{F_i'}(v) \geq 2\}$, and $F^* = F \setminus \bigcup_{i \in \ell} (F_i' \cup F_i'')$.
Also, we say that $A_i$ is \textit{final} if, for all $v \in F_i'$, $\deg(v) = 2$.

\begin{lemma}
	If there is some $A_i$ of $\varphi$ which is not final and no Stopping/ Reduction Rule is applicable, then configurations B1, or B2 exist in the partitioned graph.
\end{lemma}

\begin{proof}
	Let $v_1$ be a degree three vertex of $F_i'$, $a_i$ its neighbor in $A_i'$ and $v_2$ the other neighbor of $A_i'$ in $F$.
	We know that $vv_2 \notin E(G)$, otherwise rule R1 would be applicable.
	Now, let $v_3,v_4$ be two of the other neighbors of $v$.
	If both are in $F$, we have a configuration B1; otherwise at most one of them is not in $F \cup A_i$, say $v_3$, or we would have applied rules S2 or rule R2. This last situation implies that configuration B2 is present.
\end{proof}

We may now assume that every $A_i$ is final, and that no reduction or stopping rule is applicable.
Our goal is to show that, if none of our branching configurations exist, then $\varphi^* = (A_1 \cup F'_1 \cup F''_1, \dots, A_{\ell - 1} \cup F'_{\ell - 1} \cup F''_{\ell - 1}, A_\ell \cup F'_\ell \cup  F''_\ell \cup F^* )$ is an matching $\ell$-multicut of $G$.
Before proving that, however, we have to guarantee that the sets $F'_i, F''_i$ are a partition of $F$.

\begin{lemma}
	If there exists $i,j \in [\ell]$ with $F'_i \cap F'_j \neq \emptyset$, then rule B7 is applicable.
\end{lemma}

\begin{proof}
	Let $v_1 \in F'_i \cap F'_j$; since $A'_i$ and $A'_j$ are final, $\deg(v_1) = 2$ and its two neighbors, $a_i, a_j$, have one extra neighbor each, say $v_2$ and $v_3$.
	If $v_2 = v_3$, however, $v_2 \in F'_i$, and has degree equal to two; but this implies that $N(v_1) = N(v_2) = \{a_i, a_j\}$, and rule R4 could have been applied.
	All that remains now is the case where $v_2 \neq v_3$, but this is precisely configuration B7, as desired.
\end{proof}

\begin{lemma}
	If $A_i$ and $A_j$ are final, $F'_i \cap F''_j = \emptyset$.
\end{lemma}

\begin{proof}
	Suppose that there is some $v \in F'_i \cap F''_j$.
	By the definitions of $F_i'$ and $F''_j$, $v$ has degree two, one neighbor in $A_i$, and \textit{two} neighbors in $F'_j$, a contradiction.
\end{proof}

\begin{lemma}
	If there exists $i,j \in [\ell]$ with $F''_i \cap F''_j \neq \emptyset$, rule B6 is applicable.	
\end{lemma}

\begin{proof}
	Let $v_1$ a vertex of $F''_i \cap F''_j$.
	By the previous lemma and the definition of $F''_i$, we can readily observe that $v_1$ has four distinct neighbors: $v_2, v_3, v_4, v_5$, such that $v_2, v_3 \in F'_i$ and $v_4, v_5 \in F'_j$.
	Let $a_i$ be the neighbor of $v_2$ in $A_i$, $v_2'$ the other neighbor of $a_i$ in $F$.
	Define $a_i'$ and $v_3'$ similarly for $v_3$; $a_j$ and $v_4'$ for $v_4$; and $a_j'$ and $v_5'$ for $v_5$.
        Note that $v_2' \neq v_3$, otherwise we could apply rule R5; moreover, $v'_2 \neq v'_3$ (resp. $v'_4 \neq v'_5$), or rule R2 would be applicable.
        A similar analysis holds for $v_3,v_3',v_4, v_4'$.
	Consequently, $\{v_1, v_2, a_i, v_2', v_3, a'_i, v'_3, v_4, a_j, v'_4, v_5, a'_j, v'_5\}$ form configuration B6.
\end{proof}

These last few results prove that not only we have a partition of $F$, but that $\varphi^*$ is a partition of $V(G)$ since the $A_i's$ are disjoint by construction.
Define $A_i^* = A_i \cup F'_i \cup F''_i$ and $A_\ell^* = A_\ell \cup F'_\ell \cup F''_\ell \cup F^*$.
What remains to be shown is that it is, in fact, a matching $\ell$-multicut of $G$. In the following, we always assume that no stopping or reduction rules are applicable.

\begin{lemma}
	If no more branching rules are applicable, then for every $i$ and every $v \in A_i$, $\deg_{V(G) \setminus A_i^*}(v) \leq 1$. 
\end{lemma}

\begin{proof}
	First, $v$ has at most one neighbor in $\bigcup_ {j \neq i} A_j$, otherwise rule S4 would have stopped the algorithm.
	For the case where $v$ has one neighbor in $w \in A_j$ and one neighbor $u \in F$, first note that $wu \notin E(G)$, as rule S3 is not applicable; moreover, by rule R3, $u$ must have been added to $A_i$, and so no such $u$ exists.
	Thus, the only possibility is that $v$ has more than one neighbor in $F$ and none in $\bigcup_ {j \neq i} A_j$, implying $N_F(v) \subseteq F'_i$, but $F'_i \subseteq A_i^*$ and the claim follows.
\end{proof}

\begin{lemma}
	If no more branching rules are applicable, then for every $i$ and every $v \in F'_i$, $\deg_{V(G) \setminus A_i^*}(v) \leq 1$. 
\end{lemma}

\begin{proof}
	This follows immediately from the hypothesis that $A_i$ is final, i.e. that for every $v \in F'_ i$ we have $\deg(v) = 2$ and $N(v) \cap A_i \neq \emptyset$.
\end{proof}

\begin{lemma}
	If no more branching rules are applicable, then for every $i$ and every $v \in F''_i$, $\deg_{V(G) \setminus A_i^*}(v) \leq 1$. 
\end{lemma}

\begin{proof}
	If $v$ has a neighbor in $A_j$, rule B5 is applicable, since $v \in F''_i$.
	On the other hand, if $v$ has a neighbor in $F$, we can apply rule B4' with $v = v_1$.
        Note that the graph must be organized as the mentioned rules, otherwise we would have been able to apply rule R5 to the common neighbors of $v$ and $A_i$ in $F'_i$.
\end{proof}

\begin{lemma}
	If no more branching rules are applicable, then for every $v \in F^*$, $\deg_{V(G) \setminus A^*_\ell}(v) \leq 1$. 
\end{lemma}

\begin{proof}
	We know that $v$ does not have two neighbors in some $A_j$, but it could be the case that $v$ has neighbors $a_i \in A_i$, $a_j \in A_j$.
	Moreover, note that $x \in N(v) \cap A_i$ cannot have a second neighbor in $F$, otherwise we would have $v \in F'_i$, and the same holds for $y \in N(v) \cap A_j$ and $F'_j$.
	As such, if $\deg(v) = 2$, we can still apply rule R6.
	Otherwise, if $\deg(v) \geq 3$, configuration B3 shows up with $v = v_1$; note that we cannot have a third $A_h$ in $N(v)$, otherwise rule S2 would have been applicable.
	This allows us to conclude that $v$ does not have a neighbor in more than one $A_i$.
	Suppose now that $u \in F \setminus F^*$ is a neighbor of $v$, and $a_j \in A_j \cap N(v)$.
	If $u \in F'_i$ ($i$ may be equal to $j$), it follows that rule B8 is applicable with $v = v_3$ and $u = v_1$.
	If, on the other hand, $u \in F''_i$, we have configuration B4' where $u = v_1$ and $v = v_4$.
	Consequently, $v$ has no neighbor in $A_j$, for $j \neq \ell$.
	
	Now, it must be the case that both neighbors $x,y$ of $v$ are in $F \setminus F^*$.
	Note that $\{x,y\} \nsubseteq F'_i$, otherwise $v \in F''_i$.
	For now, suppose that $x \in F'_i$ and $y \in F'_j$.
	If $\deg(v) = 2$, rule R7 may be applied (with $u$ = $v$); so $\deg(v) \geq 3$ and we have configuration B4, again, with $v = v_1$.
	Suppose, then that $x$ is actually in $F''_i$; by the exact same argument, it holds that B4' is applicable with $v = v_4$ and $x = v_1$.
	The case where $x$ and $y$ are in $F''_i$ and $F''_j$, respectively, is identical.
\end{proof}

\begin{theorem}
	If no Stopping, Reduction, or Branching rule is applicable, $\varphi^*$ is a matching $\ell$-multicut of $G$.
    Moreover, \pname{Matching Multicut} can be solved in $\bigOs{\alpha_\ell^n}$, with $\alpha_\ell \leq \sqrt{\ell}$ for a graph on $n$ vertices.
\end{theorem}

%% file: treewidth.tex
\section{FPT Algorithm by Treewidth}

Let $(T, \{X_t\}_{t \in V(T)})$ be a nice tree decomposition of a graph $G$ with $n$ vertices, with $T$ corresponding to the tree of the nice tree decomposition and $X_t$ being the bag corresponding to vertex $t$. Suppose $T$ is rooted at a vertex $\roott$, that $X_{\roott} = \emptyset$. Let $V_t$ be the union of all the bags present in the subtree rooted at $t$. Finally, define $G_t = G[V_t]$.

Our goal is to have $c[t, \mathcal{P}, \Ext] = \ell$ if and only if $\ell$ is the maximum integer such that $(G_t, \ell)$ is a \YES\ instance of \pname{Matching Multicut} that respects $\mathcal{P}$ and $\Ext$, which we now formally define. First, $\mathcal{P}$ is a function $\mathcal{P}: X_t \mapsto X_t $ with $\mathcal{P}(v)$ corresponding to the vertex in $X_t$ with the smallest label that is present in the same set as $v$ in the partition. In other words, $\mathcal{P}$ is responsible for representing which partition of $V_t$ we have assigned each vertex of $X_t$ to.
Note that $\mathcal{P}(v) \leq v$.
Finally, $\Ext : X_t \longrightarrow \{0, 1\}$ is a function that signals whether each vertex in $X_t$ has a neighbor in a different set in the partition.
We denote by $\mathcal{P}|_{\overline{v}}$  the restriction of $\mathcal{P}$ to $X_t \setminus \{v\}$. Note that we can easily update each value in $\mathcal{P}|_{\overline{v}}$ to account for the missing vertex: we pick the minimum element in $\mathcal{P}^{-1}(v) \setminus \{v\}$ and set it as the new root of the component previously identified by $v$.

We say that $c[t, \mathcal{P}, \Ext]$ is invalid if there exists a $u \in X_t$ such that $u$ has more than $\Ext(u)$ neighbors in different sets in $X_t$. We also have an invalid state if there exists a $v$ with $\mathcal{P}(v) > v$. In these cases, we define $c[t, \mathcal{P}, \Ext] = -\infty$.

Now we will show how to compute $c[ \cdot, \cdot, \cdot]$.
To simplify the analysis of the recurrences, assume that $c[t, \mathcal{P}, Ext]$ is not invalid.

\subsection{Leaf}

If $t$ is a leaf, the only possible state is $c[t, \emptyset, \emptyset] = 0$, where $\emptyset$ is used to denote functions form the empty set to itself.

\subsection{Introduce node}

If $t$ is an introduce node, let $t'$ be the child of $t$ such that $X_t = X_{t'} \cup \{v\}$ for some $v \in V(G)$. We claim that the following formula is valid:

\begin{center}
    $c[t, \mathcal{P}, \Ext] = \begin{cases} 
    c[t', \mathcal{P}|_{\overline{v}}, \Ext'] + 1 & \mathcal{P}(u) = v \iff u = v \\
    c[t', \mathcal{P}|_{\overline{v}}, \Ext'] & \text{otherwise} 
    \end{cases}$
\end{center}

First, we define $\Ext'$. If $\Ext(v) = 0$, then $\Ext'(u) = \Ext(u)$ for all $u \in X_{t'}$. If $\Ext(v) = 1$, by the construction of the nice tree decomposition, the neighbor $w$ of $v$ in another set is in $X_{t'}$. Thus, we set $Ext'(w) = 0$ and $Ext'(u) = Ext(u)$ for the remaining vertices in $X_{t'}$. 

Let $\mathcal{A} = \{A_1, \dots, A_{\ell}\}$ be a solution to $G_t$ respecting $\mathcal{P}$ and $\Ext$, i.e. $c[t, \mathcal{P}, \Ext] = \ell$.
If w.l.o.g. $\{v\} = A_1$, then $\mathcal{P}(u) = v$ if and only if $u = v$; moreover, $\mathcal{A} \setminus \{A_1\}$ is a partition of $G_{t'}$ of size $\ell -1$,constrained by $\mathcal{P}|_{\overline{v}}$ and $\Ext'$, which is given in the first case of the recurrence.
Note that there can be at most one $u \in X_t \cap N(v)$ with $\Ext(u) = 1$ since $c[t, \mathcal{P}, \Ext]$ is valid.
If, however, $A_1 \setminus \{v\} \neq \emptyset$, then it follows that $\{A_1 \setminus \{v\}, \dots, A_\ell\}$ is a partition of $G_{t'}$ that satisfies $\mathcal{P}|_{\overline{v}}$ and $\Ext'$: we have at most elected a new root for $A_1$ and changed the entry $\Ext$ of at most one other $u \in X_{t'} \cap X_t \cap N(v)$.
Once again, this is correctly captured in the second case of the recurrence.

Now, let $\mathcal{A} = \{A_1, \dots, A_{\ell}\}$ be a solution to $G_{t'}$ represented by $\mathcal{P}'$  and $\Ext'$.
If $\mathcal{A} \cup \{\{v\}\}$ is a matching multicut of $G_t$ on $\ell+1$ parts, then this is correctly captured in the first case of the above recurrence: we elect no new representative for the elements of $\mathcal{A}$, set $\mathcal{P}(v) = v$ and set $\Ext(v) = \Ext(u) = 1$ if and only if $u$ is the unique neighbor of $v$ in $X_{t'}$, otherwise $\Ext(u) = \Ext'(u)$ and, if $v$ has no neighbors, $\Ext(v) = 0$.
If there is some $A_i$ where $\mathcal{A} \setminus \{A_i\} \cup \{A_i \cup \{v\}\}$ is a matching multicut of $G_t$, then this is captured in the second case of the recurrence: it suffices to observe that we may have to set $\mathcal{P}(u) = v$ for all $u \in A_i \cap X_t$, $\mathcal{P}(w) = \mathcal{P}'(w)$ for all other vertices of $X_t$, and that at most one element $u$ in $A_j \neq A_i$ must have its $\Ext'$ value updated to $1$, and even then only if $u \in N(v)$.

In terms of complexity, note that both $\Ext'$ and $\mathcal{P}|_{\overline{v}}$ can be computed in $\bigO{|X_t|}$ time, and so each entry $c[t, \mathcal{P}, \Ext]$ of an introduce node $t$ can be computed in $\bigO{|X_t|}$ time.

\subsection{Join node}

If $t$ is a join node with children $t_1$ and $t_2$, we say that the constraints $\Ext_1, \Ext_2$ with respect to $t_1, t_2$ respectively are consistent with the constraints $\mathcal{P}$ and $\Ext$ with respect to $t$ if the following conditions hold: 
\begin{enumerate}
    \label{cond:1} \item if $\Ext(u) = 0$, then $\Ext_1(u) = \Ext_2(u) = 0$;
    \label{cond:2} \item if $\Ext(u) = 1$ and $u$ has a neighbor in a different set in $X_t$, then $\Ext_1(u) = \Ext_2(u) = 1$;
    \label{cond:3} \item if $\Ext(u) = 1$ and $u$ does not have a neighbor in a different set in $X_t$, then $\Ext_1(u) + \Ext_2(u) = 1$.
\end{enumerate}

We say that a function $f$ is compatible with a partition of vertices $\mathcal{A}$ if for every $u \in \mathcal{A}$ it holds that $u$ has $f(u)$ neighbors in a different set with respect to $\mathcal{A}$.

With this, we assert that the following formula is valid:
\begin{center}
    $c[t, \mathcal{P}, \Ext] = \max\limits_{\Ext_1, \Ext_2} c[t_1, \mathcal{P}, \Ext_1] + c[t_2, \mathcal{P}, \Ext_2] - x$,
\end{center}
where $x$ is the number of distinct elements in $\mathcal{P}$ and the maximum is taken over all pairs $\Ext_1, \Ext_2$ consistent with the constraints $\mathcal{P}, \Ext$.

Let $\mathcal{A} = \{A_1, \ldots, A_\ell \}$ be a solution to $G_t$ respecting $\mathcal{P}$ and $\Ext$. Then, let $\mathcal{B} = \{B_1, \ldots, B_{\ell_1}\}, \mathcal{C} = \{C_1, \ldots, C_{\ell_2}\}$ be the partitions $\mathcal{A}$ induces in $G_{t_1}$ and $G_{t_2}$ respectively.
We now claim that $\Ext_1, \Ext_2$ are compatible with $\mathcal{B}, \mathcal{C}$. As these partitions were obtained by erasing elements from subsets of $\mathcal{A}$, condition 1 is satisfied. Moreover, as $X_t \subseteq G_{t_1}, G_{t_2}$, condition 2 is also satisfied. Now, consider that $u \in X_t$ satisfies condition 3 and its neighbor from a different set is $w$. Then $w \in G_t \setminus X_t$, implying that $w$ must belong to exactly one of $G_{t_1}, G_{t_2}$ and exactly one of $\Ext_1(u), \Ext_2(u)$ equals 1. Thus, condition 3 is satisfied as well. Finally, we have that $\ell_1 + \ell_2  = \ell + x$, as the parts of $\mathcal{B}, \mathcal{C}$ containing vertices in $X_t$ are the only ones accounted for twice when summing $\ell_1 + \ell_2$, as all other parts were already forgotten either in $G_{t_1}$ or $G_{t_2}$.

In the other direction, let $\mathcal{B} = \{B_1, \ldots, B_{\ell_1}\}$ be a solution of $G_{t_1}$ respecting $\mathcal{P}, \Ext_1$ and $\mathcal{C} = \{C_1, \ldots, C_{\ell_2}\}$ be a solution of $G_{t_2}$ respecting $\mathcal{P}, \Ext_2$. Moreover, assume that $\Ext_1, \Ext_2$ are consistent with $\mathcal{P}$ and $\Ext$. W.l.o.g, notice that only the sets $B_1, \ldots, B_x$ and $C_1, \ldots, C_x$ contain vertices from $X_t$. If $v_1, \ldots, v_x \in X_t$ are the representatives in $\mathcal{P}$ (i.e. $\mathcal{P}(v_i) = v_i$), then assume that $v_i \subseteq B_i, C_i$ for $1 \leq i \leq x$. 

We claim that the partition $\mathcal{A} = \{B_1 \cup C_1, \ldots, B_x \cup C_x, B_{x+1}, \ldots, B_{\ell_1}, C_{x+1}, \\ \ldots, C_{\ell_2}\}$ is a solution of size $\ell_1+\ell_2-x$ to $G_t$.
Observe that vertices that are in the same set in $\mathcal{B}$ or $\mathcal{C}$ continue to be in the same set in $\mathcal{A}$. Thus, as vertices from $G_t \setminus X_t$ have the same set of neighbors in $G_{t_1}$ (or $G_{t_2}$) and in $G_t$, they must have at most one neighbor from a different set in $\mathcal{A}$. 

Now, notice that for every node $v \in X_t$ it holds that $v$ has at most $\Ext_1(v) + \Ext_2(v)$ neighbors of different sets in $\mathcal{A}$. If some vertex $u \in X_t$ have two neighbors of different sets in $\mathcal{A}$, then $\Ext(u)$ clearly cannot fall into condition $\hyperref[cond:1]{1}$ of consistency. The same holds for condition $\hyperref[cond:2]{2}$, as the neighbor of $u$ in different set in $\mathcal{B}, \mathcal{C}$ would be the same. Finally, $\Ext(u)$ must not satisfy condition $\hyperref[cond:3]{3}$ either, as this implies that either $\Ext_1(u) = 0$ or $\Ext_2(u)=0$. For the last two cases, notice that $u$ has exactly one neighbor of different set in $\mathcal{A}$, and so $\mathcal{A}$ is a solution that respects both $\mathcal{P}$ and $\Ext$.

Unlike most dynamic programming algorithms, our join nodes are effectively quite cheap. Our only real choice is represented by condition 3: for each $u \in X_t$ with a crossing neighbor outside of $X_t$, we must check whether this neighbor is in $G_{t_1}$ or $G_{t_2}$, so we effectively have at most $2^{|X_t|}$ possibilities to investigate per entry of a join node.

\subsection{Forget node}

If $t$ is a forget node with child $t'$, then $X_t = X_{t'} \setminus \{v\}$ for some $v \in X_{t'}$. Furthermore, it is valid that $G_t = G_{t'}$.

To calculate $c[t, \mathcal{P}, \Ext]$, we are interested in all partitions $\mathcal{P}'$ such that $\mathcal{P'}|_{\overline{v}} = \mathcal{P}$. We call a $\mathcal{P}'$ that satisfies this property an augmentation of $\mathcal{P}$. Then, the following formula is valid:

\begin{center}
$c[t, \mathcal{P}, \Ext] = \max\limits_{\mathcal{P}'} c[t', \mathcal{P}', \Ext]$,
\end{center}
where the maximum runs over all $\bigO{|X_t|}$ augmentations of $\mathcal{P}$.

Now we show that the formula is valid. Let $\mathcal{A}$ be a partition of $G_t$ respecting $\mathcal{P}, \Ext$. As $\mathcal{A}$ induced by the vertices in $X_t$ is the partition represented by $\mathcal{P}$ and $X_{t'} = X_t \cup v$, it holds that $\mathcal{A}$ induced by the vertices in $X_{t'}$ is the partition represented by some augmentation of $\mathcal{P}$. Therefore $c[t, \mathcal{P}, \Ext] \leq \max\limits_{\mathcal{P}'} c[t', \mathcal{P}', \Ext]$.

In the other direction, let $\mathcal{A}$ be a valid partition of $G_t$ respecting $\mathcal{P}', \Ext$ for some augmented partition $\mathcal{P}'$. Thus, by definition $\mathcal{P'}|_{\overline{v}} = \mathcal{P}$ and $\mathcal{A}$ induced by the vertices in $X_t$ is the partition represented by $\mathcal{P}$. Then $\mathcal{A}$ also respects $\mathcal{P}, \Ext$ and we conclude that $c[t, \mathcal{P}, \Ext] \geq \max\limits_{\mathcal{P}'} c[t', \mathcal{P}', \Ext]$.

\begin{theorem}
    If given a nice tree decomposition of width $k$ of the $n$-vertex graph $G$, there exists an algorithm that solves \pname{Matching MultiCut} in $2^{k\log k}n$ time.
\end{theorem}

\begin{proof}
    The correctness follows immediately from the previous analysis and recurrences for each node type: as we compute the maximum number of parts $c[\roott, \emptyset, \emptyset] = r$ a matching multicut can have in $G$, answering if $(G,\ell)$ is a \YES\ instance is simply answering if $r \geq \ell$.
    
    In terms of complexity, our most expensive nodes are join nodes.
    We can compute each entry $c[t,\mathcal{P}, \Ext]$ of this node type in $\bigO{2^kk}$ time and, as we have no more than $nk \times k^k \times 2^k$ table entries to compute, it follows immediately that our algorithm runs in $2^{\bigO{k\log k}}n$ time.
\end{proof}

%% file: enum_vc.tex
\subsection{Vertex Cover}
\label{sec:enum_vc}

In this section, we consider the parameterization of the matching multicut problem by the vertex cover number of the input graph. This parameterization of \pname{Enum Matching Cut} was previously studied in~\cite{golovach2022refined}.
We show that the enumeration kernel constructed by the authors of~\cite{golovach2022refined} is also an enumeration kernel for \pname{Enum Matching Multicut}.

Computing a minimum vertex cover $\tau(G)$ of $G$ is well known to be \NPH, but one can find a 2-approximation by taking the end-vertices of a maximal matching of $G$. Notice also that for every matching multicut $M \subseteq E(G)$, $|M| \leq \tau(G)$ since $M$ is a matching.
Throughout this section, we assume that the vertex cover $X$ of size $k \leq 2\tau(G)$ is given together with the input graph.

We describe the kernel constructed in \cite{golovach2022refined}.
Assume for simplicity that $G$ contains no isolated vertices. Let $I = V(G) \setminus X$. Recall that $I$ is an independent set. Denote by $I_1$ and $I_{\geq 2}$ the subsets of vertices of $I$ with degree $1$ and at least $2$, respectively. We use the following marking procedure to label some vertices of $I$.

\begin{enumerate}
    \item[(i)] For every $x \in X$, mark an arbitrary vertex of $N(x) \cap I_1$ (if it exists).
    \item[(ii)] For every two distinct vertices $x, y \in X$, select an arbitrary set of $\min \{3, |N(x) \cap N(y) \cap I_{\geq 2}|\}$ vertices in $I_{\geq 2}$ that are adjacent to both $x$ and $y$, and mark them for the pair $\{x, y\}$.
\end{enumerate}

Denote by $Z$ the set of marked vertices of $I$. Define $H = G[X \cup Z]$. Notice that $|V(H)| \leq 2|X| + 3\binom{|X|}{2} = \bigO{k^2}$. This completes the description of the basic compression algorithm that returns $H$. The key property of $H$ is that it keeps all matching cuts of $G' = G - I_1$, including all matching multicuts of $G'$. Formally, we define $H' = H - I_1$ and show the following lemma.

\begin{lemma}
    A set of edges $M \subseteq E(G')$ is a matching $t$-multicut of $G'$ if and only if $M \subseteq E(H')$ and $M$ is a matching $t$-multicut of $H'$ with as many parts.
\end{lemma}

\begin{proof}
    Suppose that $M \subseteq E(G')$ is a matching $t$-multicut of $G'$ and assume that $M = E_{G'}(A_1, \dots, A_t)$ for a partition $\{A_1, \dots, A_t\}$ of $V(G')$. As $M$ is a matching $t$-multicut, $E_{G'}(A_i, V(G') \setminus A_i)$ is a matching cut of $G'$. It follows from \cite[Lemma 10]{golovach2022refined} that $E_{G'}(A_i, V(G') \setminus A_i) \subseteq E(H')$. As this holds for every $i \in [t]$, we have that $M$ is a matching $t$-multicut of $H'$.
    
    Notice that $M$ is a matching $t$-multicut for a partition $\{A_1, \dots, A_t\}$ if and only if the partitions $\{A_i, V \setminus A_i\}$ define matching cuts. Therefore, the opposite direction follows from the same argument.
\end{proof}

To see the relationship between matching multicuts of $G$ and $H$, we define a special equivalence relation for the subsets of edges of $G$. For a vertex $x \in X$, let $L_x = \{xy \in E(G) \mid y \in I_1\}$, that is, $L_x$ is the set of pendant edges of $G$ with exactly one end-vertex in the vertex cover. Observe that if $L_x \neq \emptyset$, then there is $l_x \in L_x$ such that $l_x \in E(H)$, because for every $x \in X$, a neighbor in $I_1$ is marked if it exists. We define $L = \bigcup_{x \in X} L_x$. Notice that each matching multicut of $G$ contains at most one edge of every $L_x$. We say that two sets of edges $M_1$ and $M_2$ are equivalent if $M_1 \setminus L = M_2 \setminus L$ and for every $x \in X$, $|M_1 \cap L_x| = |M_2 \cap L_x|$. It is straightforward to verify that the introduced relation is indeed an equivalence relation. It is also easy to see that if $M$ is a matching multicut of $G$, then every $M' \subseteq E(G)$ equivalent to $M$ is a matching multicut. The next lemma follows directly from \cite[Lemma 10]{golovach2022refined}.

\begin{lemma} \label{lem:vc_equiv}
A set of edges $M \subseteq E(G)$ is a matching multicut of $G$ if and only if $H$ has a matching multicut $M'$
equivalent to $M$.
\end{lemma}

Since Lemma \ref{lem:vc_equiv} holds for the same equivalence relation as in \cite{golovach2022refined}, the algorithm used to enumerate equivalent solutions of matching cut can also be used to enumerate equivalent solutions of matching multicut, without modifications.

\begin{theorem} \label{thm:enum_vc}
    \pname{Enum Matching Multicut} admits a polynomial-delay enumeration kernel with $\bigO{k^2}$ vertices when parameterized by the vertex cover number $k$ of the input graph.
\end{theorem}

By Theorem \ref{thm:enum_vc}, we have that matching multicuts can be listed with delay $k^{\bigO{k^2}} \cdot n^{\bigO{1}}$. We believe that this running time can be improved and the dependence on the vertex cover number can be made single exponential.

%% file: enum_dcc.tex
\subsection{Distance to Co-cluster}

A 3-approximation for this parameter can easily computed in polynomial time: for every induced $\overline{P_3}$, add all three of its vertices to the modulator.
As such we assume that, along with $(G, \ell)$, we are given a set $S$ of size $k \leq 3\dcc(G)$ so that $G \setminus S$ is a co-cluster graph. We break down our analysis in three cases: if $G \setminus S$ has at least three parts, two large parts, or neither of the previous two.

\subsubsection{More than three parts}

\begin{lemma}
    If $G \setminus S$ has at least $3$ disjoint independent sets, then $G \setminus S$ is an indivisible set of vertices.
\end{lemma}

\begin{proof}
    It suffices to note that any three vertices of $G \setminus S$ that do not belong to the same independent set are pairwise adjacent and, consequently, form a triangle, which is indivisible.
    Since indivisibility is transitive, the entirety of $G \setminus S$ is indivisible.
\end{proof}

With that, suppose that $G \setminus S$ is a complete multipartite graph that can be partitioned into at least three independent sets. We can handle the cases where $G \setminus S$ has at most two independent sets using the FPT delay algorithm parameterized by distance to cluster previously discussed.

Therefore, from now on, we assume that no edge of $G \setminus S$ can belong to an $\ell$-Matching Multi-Cut.

We will apply the following rules exhaustively and in order:

\begin{rrule} \label{red:10} If there exists $u \in S$ with two neighbors $v, w \in V(G) \setminus S$, add the edge $uz$ to $E(G)$ for every $z \in V(G) \setminus S$ and then remove $u$ from $S$.
\end{rrule}

\begin{rrule} \label{red:11} If there exists $u \in V(G) \setminus S$ without a neighbor in $S$ and $(G \setminus S) \setminus \{u\}$ has at least three disjoint independents sets, remove $u$ from $G$.
\end{rrule}

Let $H$ be the graph obtained after applying the above rules exhaustively. Note that after each application of \ref{red:10}, it still holds that no edge in $G \setminus S$ belongs to an $\ell$-MMC. Let $S_G, S_H$ be the sets $S$ obtained before and after applying the rules, respectively, and $R$ the set of vertices removed by \ref{red:11}.

\begin{lemma} \label{lem:16}
    It holds that $|V(H)| \leq 2k$.
\end{lemma}

\begin{proof}
    Since $S_H \subseteq S_G$ and $|S_G| = k$, $|S_H| \leq k$. If there were $k+1$ vertices in $H \setminus S_H$, at least one vertex in $S_H$ would have two neighbors in $H \setminus S_H$, which is absurd because rule \ref{red:10} is not applicable on $H$.
\end{proof}

\begin{lemma}\label{lem:dcc3}
    A set of edges $M$ is a matching $\ell$-multicut of $H$ if and only if $M$ is a matching $\ell$-multicut of $G$.
\end{lemma} 

\begin{proof}
It suffices to note that every edge in $E(G) \setminus E(H)$ was removed due to Rule~\ref{red:10} and all edges in $E(H) \setminus E(G)$ were added by Rule~\ref{red:11}.
In either case, all of these edges are interior to the co-cluster graph obtained by the removal of $S_H$ or $S_G$, which is indivisible and thus must be always entirely contained a single part of a matching multicut of both $H$ and $G$.
As such, every edge set that forms a matching $\ell$-multicut also forms a matching $\ell$-multicut in the other.
\end{proof}

\subsubsection{Two large parts}
For this case, suppose that $G \setminus S$ is isomorphic to the complete bipartite graph $K_{a,b}$, with $a \leq b$, $a \geq 2$ and $b \geq 3$.
Note that $G \setminus S$ is also an indivisible subgraph of $G$, and we can proceed as in the previous case, but instead of rules~\ref{red:10} and~\ref{red:11}, we apply the following two, which play the exact same role in our current situation.

\begin{rrule}\label{rrule:dcc2-1}
    If there exists $u \in S$ with at least two neighbors in $V(G) \setminus S$, then erase all edges from $u$ to $G \setminus S$ and add all edges between $u$ and the smallest part of $G \setminus S$.
\end{rrule}

\begin{rrule}\label{rrule:dcc2-2}
    If there exists $u \in V(G) \setminus S$ with no neighbors in $S$ and both parts of $G \setminus S$ have at least 3 vertices, then remove $u$ from $G$.
\end{rrule}

With this, we can prove analogous results to Lemmas~\ref{lem:16} and~\ref{lem:dcc3}, which we omit the proofs for brevity.

\begin{lemma} \label{lem:16-2}
    If rules~\ref{rrule:dcc2-1} and~\ref{rrule:dcc2-2} are not applicable, then the reduced graph $H$ satisfies $V(H) \leq 2k + 2$.
\end{lemma}

\begin{lemma}
    A set of edges $M$ of $H$ is a matching $\ell$-multicut if and only if $M$ is also a matching $\ell$-multicut of $G$.
\end{lemma}

\subsubsection{Otherwise}
For our final cases, note that $G \setminus S$ has either one or two parts, and the latter case it must hold that $a \leq 2$. As such, if we do have two parts, we can augment $S$ with the at most two elements of the smallest part and have a vertex cover of size at most $|S| + 2$.
As such, we now have a vertex cover of $G$, and we apply the kernelization algorithm of section~\ref{sec:enum_vc}.
Combining Lemmas~\ref{lem:16} and~\ref{lem:16-2} and Theorem~\ref{thm:enum_vc}, we obtain our main kernelization result:

\begin{theorem} \label{thm:enum_dcc}
    \pname{Enum Matching Multicut} admits a polynomial-delay enumeration kernel with $\bigO{k^2}$ vertices when parameterized by the vertex-deletion distance to co-cluster $k$ of the input graph.
\end{theorem}

%% file: enum_dc.tex
\subsection{Distance to Cluster}

In this section, we present a \DFPT\ enumeration algorithm for \pname{Enum Matching Multicut}, parameterized by the vertex-deletion distance to cluster.
We base our result on the quadratic kernel \pname{Matching Cut} given in~\cite{matching_cut_ipec}. The authors apply several reduction rules until they reach a kernel of size \( \text{dc}(G)^{\mathcal{O}(1)} \).
We use a subset of these rules as a starting point for our enumeration algorithm, then expand them a more careful analysis and needed technicalities for an enumeration algorithm. As such, our goal is to prove the following theorem.

\begin{theorem} \label{prop:1}
There is an algorithm for \pname{Enum Matching Multicut} on $n$-vertex graphs with distance to cluster \( \dc(G) \leq t \) of delay \( 2^{\mathcal{O}(t^3 \log t)} + n^{\mathcal{O}(1)}\).
\end{theorem}

Our strategy to enumerate all possible matching multicuts can be divided into 5 steps:

\begin{enumerate}
    \item We apply reduction rules, similar to the kernelization steps described in \cite{matching_cut_ipec}. This step is polynomial in $|G|$. \label{step1}
    \item We enumerate all possible matching multicuts of a smaller instance of size $\mathcal{O}(t^3)$. This step takes a total time of $2^{\mathcal{O}(t^3 \log t)}$.\label{step2}
    \item Given a matching multicut generated in step 2, we create an instance of \pname{Enum Set Packing}, where the ground set has size $t$ and the number of sets is potentially $2^t$. All solutions are enumerated in total time $2^{\mathcal{O}(t^2)}$, and then each solution is extended to form a matching multicut.\label{step3}
    \item Given a matching multicut from step 3, we increase the number of partitions by considering clusters of size 2 with only one edge to $U$. At this step, we ensure to have at least $\ell$ partitions. As the number of matching multicuts with at least $\ell$ parts can be exponential in $\ell$, we start to be concerned with the delay of the enumeration and no longer with the total time.\label{step4}
    \item As a final step, we lift the previous solutions and obtain correct solutions for the original instance.\label{step5}
\end{enumerate}

\subsubsection{Step 1 - Reduction Rules}

We proceed with \textbf{Step}~\ref{step1} by first applying the first 7 reduction rules described in \cite{matching_cut_ipec}, except by the first one, as we want to enumerate matching multicuts and not only give an answer for the decision problem. For completeness, we explicitly write all reduction rules used in our algorithm.

Let $G$ be our input graph and $U \subseteq V(G)$.
Then we call $U$ \textit{monochromatic} in $G$ if for every matching multicut $\{A_1, \dots, A_k\}$ of $G$, $U \subseteq A_i$ for some $1 \leq i \leq k$.
Let $U = \{u_1, \dots, u_{|U|}\}$ denote a vertex set such that $G \setminus U$ is a cluster graph.
Computing $\text{dc}(G)$ is NP-hard, but one can find a 3-approximation by taking a maximal set of vertex disjoint induced paths with 3 vertices, as no cluster graph contains an induced $P_3$.
During the algorithm, we maintain a partition of $U$ into $U_1, \dots, U_r$ such that each $U_i$ is monochromatic.
The initial partition contains one set for each vertex of $U$, that is, $U_i := {u_i}, i \in [|U|]$. We call the sets of the partition the monochromatic parts of $U$.
During this step we may merge two sets $U_i$ and $U_j$, $i \neq j$, which is to remove $U_i$ and $U_j$ from the partition and to add $U_i \cup U_j$.
We say that merging $U_i$ and $U_j$ is safe if $U_i \cup U_j$ is monochromatic in $G$.

To identify clusters that form monochromatic sets together with some monochromatic parts of $U$, we use the following notation. For each monochromatic part $U_i$ of $U$, let $N^*(U_i)$ (denoted by $N^2(U_i)$ at \cite{matching_cut_ipec}) denote the set of vertices $v \in V \setminus U$ such that at least one of the following holds:

\begin{itemize}
    \item $v$ has two neighbors in $U_i$,
    \item $v$ is in a cluster of size at least three in $G \setminus U$ that contains a vertex that has two neighbors in $U_i$, or
    \item $v$ is in a cluster $C$ in $G[V \setminus U]$ and some vertex in $U_i$ has two neighbors in $C$.
\end{itemize}

It is clear that $U_i \cup N^*(U_i)$ is monochromatic.  

Unless stated otherwise, the following reduction rules are lifted straight from \cite{matching_cut_ipec}, and so we direct the reader to it for the corresponding safeness proofs.

\begin{rrule} \label{red:1}
    If there is a vertex $v$ that is contained in $N^*(U_i)$ and $N^*(U_j)$ for $i \neq j$, then merge $U_i$ and $U_j$. 
\end{rrule}

\begin{rrule} \label{red:2}
    If there are three vertices $v_1, v_2, v_3$ in $V$ that have two common neighbors $u \in U_i \text{ and } u' \in U_j$, $i \neq j$, then merge $U_i$ and $U_j$.
\end{rrule}

In the following, a cluster consisting of two vertices is an \textit{edge cluster}, all other clusters are \textit{nonedge clusters}. We show that there is a bounded number of nonedge clusters that are not contained in some $N^*(U_i)$ and that do not form a matching with $U$. We call those clusters ambiguous. More precisely, we say that a vertex in $V \setminus U$ is ambiguous if it has neighbors in $U_i$ and $U_j$ where $i \neq j$. A cluster is ambiguous if it contains at least one ambiguous vertex. In contrast, we call a cluster fixed if it is contained in $N^*(U_i)$ for some $U_i$.

\begin{rrule} \label{red:3}
If there are two clusters $C_1$ and $C_2$ that are contained in $N^*(U_i)$, then add all edges between these clusters.
\end{rrule}

Notice that after Reduction Rule \ref{red:4}, there are at most $\mathcal{O}(|U|)$ fixed clusters. 

\begin{rrule} \label{red:4}
If there is a cluster $C$ with more than three vertices that contains a vertex $v$ with no neighbors in $U$, then remove $v$.
\end{rrule}

\begin{rrule} \label{red:5}
If there is a cluster $C$ with at least three vertices and a monochromatic set $U_i$ such that $C \subseteq N^*(U_i)$, then remove all edges between $C$ and $U_i$ from $G$, choose an arbitrary vertex $u \in U_i$ and two vertices $v_1, v_2 \in C$, and add two edges $\{u, v_1\}$ and $\{u, v_2\}$. If $|U_i| = 2$, then add an edge between $u' \in U_i \setminus \{u\}$ and $v_3 \in C \setminus \{v_1, v_2\}$. If $|U_i| > 2$, then make $U_i$ a clique.
\end{rrule}

\begin{lemma}\label{lem:orig_structure}
Let $G$ be a graph with cluster vertex deletion set $U$ that is reduced with respect to Reduction Rules~\ref{red:1}--\ref{red:5}. Then $G$ has

\begin{itemize}
    \item $\mathcal{O}(|U|^2)$ ambiguous vertices.
    \item $\mathcal{O}(|U|^2)$ nonedge clusters that are either fixed or ambiguous.
    \item Each cluster contains $\mathcal{O}(|U|)$ vertices.
\end{itemize}
\end{lemma}

We call an edge cluster $\{u, v\}$ simple if $u$ has only neighbors in some $U_i$ and $v$ has only neighbors in some $U_j$ (possibly $i = j$). Observe that the number of non-simple edge clusters is already bounded, as it contains an ambiguous vertex.

\begin{rrule} \label{red:6}
If there is a simple edge cluster $\{u, v\}$ that do not form a matching with $U$, remove $u$ and $v$ from $G$.
\end{rrule}

All the above reduction rules are used in the kernelization of matching cut described in \cite{matching_cut_ipec}. For this reason, we omit the proof of safeness of those rules. Now, we start with the differences in our algorithm. The biggest difference is that we have an extra type of cluster: the ones that form a matching. As we want to enumerate all possible matching multicuts with the appropriate number of parts, it is necessary to deal with those clusters. For this reason, let $C$ be a cluster in $V \setminus U$ such that $(C, V \setminus C)$ is a matching cut, we call $C$ a \textit{matching cluster}.

\begin{observation}
    Every nonedge cluster in $G$ is ambiguous, fixed, or matching.
\end{observation}

The following reduction rule is an extension of reduction rule \ref{red:2}. And is going to be used to bound the number of matching clusters.
\begin{rrule} \label{red:7} 
    If there are three clusters $C_1, C_2, C_3$ of size at least three such that they have two common neighbors $u \in 
 N(C_k) \cap U_i \text{ and } u' \in N(C_k) \cap U_j$, for $k \in \{1, 2, 3\}$, and with $i \neq j$, then merge $U_i$ and $U_j$.
\end{rrule} 
\begin{proof}[Proof of Safeness]
    Assume that there exists a matching multicut with $u$ and $u'$ in different parts, then, at most one of the clusters $\{C_1, C_2, C_3\}$ is in a different part than $u$ and at most one is in a different part than $u'$. This is absurd and, thus $\{u, u'\}$ is monochromatic, which makes $U_i \cup U_j$ monochromatic.
\end{proof}

\begin{rrule} \label{red:8}
Let $C_1$ be a matching cluster with $|C_1| \neq 2$ or $|N(C_1)| \neq 2$. If there exists a matching cluster $C_2$ with $|C_2| = |C_1|$ and $N(C_2) = N(C_1)$, remove $C_2$.
\end{rrule}

\begin{proof}[Proof of safeness]
    Let $C_1$ and $C_2$ be matching clusters with the properties of rule \ref{red:8}. If the edges from $(C_1, U)$ belong to the final matching multicut, no edge from $(C_2, U)$ can also belong.
\end{proof}

\begin{rrule} \label{red:9}
Let $C_1$ and $C_2$ be matching clusters with $|C_1| = |C_2| = 2$, $|N(C_1)| = 2$, and $N(C_1) = N(C_2)$. If there exists a matching cluster $C_3$ with $|C_3| = |C_1|$ and $N(C_3) = N(C_1)$, remove $C_3$.
\end{rrule}
\begin{proof}[Proof of safeness]
    Let $C_1, C_2$, and $C_3$ be matching clusters as in reduction rule \ref{red:9}. If an edge from $(C_1, U)$ and an edge from $(C_2, U)$ belong to the final matching multicut, no edge from $(C_3, U)$ can also belong.
\end{proof}

If there are two clusters $C_1$ and $C_2$ satisfying conditions of Rule \ref{red:8}, paint the vertices of $C_1$ in blue. The reason for the painting of the vertices of $C_1$ is because since we have 2 clusters with the same neighborhood, at the enumeration step we want to avoid enumerating the same instance twice.

We give an analogous of lemma \ref{lem:orig_structure}.

\begin{lemma}\label{lem:matching_structure}
Let $G$ be an instance of Matching Cut with cluster vertex deletion set $U$ that is reduced with respect to Reduction Rules~\ref{red:1}--\ref{red:8}. Then $G$ has

\begin{itemize}
    \item $\mathcal{O}(|U|^2)$ matching clusters with neighbors in $U_i$ and $U_j$, for $i \neq j$.
    \item $\mathcal{O}(2^{|U|})$ matching clusters with neighbors in a single $U_i$. 
\end{itemize}
\end{lemma}
\begin{proof}
    For the matching clusters with less than 3 vertices, the bounds follow trivially from reduction rules \ref{red:8} and \ref{red:9}. 
    
    The bound on the number of matching clusters with neighbors in $U_i$ and $U_j$ follows from \ref{red:7}: suppose there are $3 \cdot \binom{|U|}{2} = \mathcal{O}(|U|^2)$ such matching clusters, then, by the pigeonhole principle, there are $u \in U_i$ and $u' \in U_j$ with $i \neq j$ and both vertices adjacent to 3 different matching clusters. 
    
    The bound on matching clusters with neighbors in a single $U_i$ is due simply by counting the number of distinct neighborhoods and by reduction rule \ref{red:8}.
\end{proof}

\subsubsection{Steps 2 and 3 - Total Time Enumeration}

Now, we start \textbf{Step}~\ref{step2}. Apart from matching clusters $C$ whose neighborhood $N(C)$ is monochromatic, we have an instance with $\mathcal{O}(|U|^3)$ vertices. As the subgraph $G[U]$ has no particular structure, there is nothing much that we can do unless we enumerate all possible matching multicuts at this step. Let $H$ be the subgraph induced by all current structures, except by matching clusters with $N(C)$ monochromatic.

\begin{lemma}
    All matching multicuts of $H$ can be enumerated in total time $2^{\mathcal{O}(|H| \log |H|)}$.
\end{lemma}
\begin{proof}
    $H$ has at most $|H|^{|H|}$ distinct partitions and for each partition, we can check in time $\mathcal{O}(|H|)$ if it forms a matching multicut.
\end{proof}

Let $M \subseteq E(H)$ be a matching multicut of $H$, and let $\mathcal{C}$ be the matching clusters not in $H$. We start \textbf{Step} \ref{step3} and incorporate $\mathcal{C}$ into $M$, creating a collection of new matching multicuts. Our plan is to enumerate the solutions of an instance for set-packing, and for each solution, construct a new matching multicut. For each matching cluster $C \in \mathcal{C}$, we look at its neighborhood $N(C)$. Recall that $N(C) \subseteq U_i$ for some $U_i$. We say that a vertex $v \in U$ is \textit{saturated} by $M$ if $v$ is the end-point of some edge in $M$. If no vertex of $N(C)$ is saturated by $M$, we consider $N(C)$ as a set $S_C$. Let $\mathcal{S}_M$ be the collection of sets generated by the Matching Multi-Cut $M$.

\begin{lemma}
    Let $\mathcal{S}$ be a collection of sets in a ground set of size $t$ that form an instance for set-packing. All possible solutions for the set packing problem can be enumerated in time $2^{\mathcal{O}(t^2)}$.
\end{lemma}
\begin{proof}
    Each solution can be formed by at most $t$ non-intersecting sets, and we have $\sum_{i=0}^{t} \binom{|\mathcal{C}|}{i} = 2^{\mathcal{O}(t^2)}$ choices.
\end{proof}

In order to extend $M$ with the matching clusters, let $\{C_1, \dots, C_q\} \subseteq \mathcal{C}$ be a collection of clusters such that $\{S_{C_1}, \dots, S_{C_q}\}$ form a solution for the set packing problem. Notice that $$M \cup \bigcup_{i=1}^q E(C_i, H)$$ is also a matching multicut for the graph $H' = H \cup \mathcal{C}$. Therefore, all matching multicuts of $H'$ can be enumerated in time $2^{\mathcal{O}(t^3 \log t)} \cdot 2^{\mathcal{O}(t^2)} = 2^{\mathcal{O}(t^3 \log t)}.$

\subsubsection{Steps 4 and 5 - Enumerate solutions with FPT delay}

We are now ready for \textbf{step} \ref{step4}. We say that an edge cluster $\{u, v\}$ is a \textit{pendant edge cluster} if $|N(\{u, v\})| = 1$. We assume that exists $uw \in E(G)$ with $w \in U$. We give special attention to these clusters because in the enumeration of matching multicuts, they can be used to increase the number of partitions by considering the edge $uv$.

Let $\mathcal{P}$ be the collection of pendant edge clusters that were removed in reduction rule \ref{red:8}. Let $M \subseteq E$ be a matching multicut with $q$ parts enumerated by step \ref{step3}. We perform the following procedure recursively to construct matching multicuts with $\ell$ parts:


\begin{itemize}
    \item[S1] If $q + |\mathcal{P}| < \ell$, then there is no matching multicut with $\ell$ parts.
    \item[E1] If $\mathcal{P}$ is empty, enumerate $M$.
    \item[B1] Let $P \in \mathcal{P}$. If $N(P)$ is not saturated, branch by either adding $uv$ to $M$ or doing nothing. In both cases, remove $P$ from $\mathcal{P}$.
    \item[B2] Let $P \in \mathcal{P}$. If $N(P)$ is saturated, branch by doing one of the following: replace the edge of $N(P)$ in $M$ and add $uw$; add $uv$ to $M$; or do nothing. In all cases, remove $P$ from $\mathcal{P}$.
\end{itemize}

\begin{lemma}
    Step \ref{step4} enumerates matching multicuts with at least $\ell$ parts with delay $\ell^{\mathcal{O}(1)}$.
\end{lemma}

Let $M$ be a matching multicut enumerated by \textbf{Step} \ref{step4}. Now we are finally ready to enumerate matching multicuts of the original graph $G$ and proceed to \textbf{Step} \ref{step5}. Recall that an edge cluster $\{u, v\}$ is simple if $N_U(u) \subseteq U_i$ and $N_U(v) \subseteq U_j$. Notice that any simple edge cluster that is not a matching cluster was removed in Reduction Rule \ref{red:6}, we consider them again for the enumeration.

Let $\mathcal{C}$ be the collection of matching clusters $C$ such that $E(C, V\setminus C) \in M$. We consider the sets $\mathcal{C}_{2}$ and $\mathcal{C}_{\neq 2}$ of $\mathcal{C}$, where the first contains all clusters of $\mathcal{C}$ with 2 vertices and $|N(C)| = 2$, and the second contains the matching clusters with sizes other than 2. Notice that matching clusters of size 2 with $|N(C)| = 1$ are the only matching clusters in $\mathcal{C}$ not covered by $\mathcal{C}_{2}$ or by $\mathcal{C}_{\neq 2}$. We do not need to enumerate equivalent solutions with them because this is already done by \textbf{Step}  \ref{step4}. Notice also that we may have pairs of clusters $C_1$ and $C_2$ in $\mathcal{C}_2$ with $N(C) = N(C')$, because of Reduction Rule \ref{red:9}. 

\begin{observation}
Any matching cluster $C'$ with $N(C') = N(C)$ and $C' \notin \mathcal{C}$ was erased by rules \ref{red:8} and \ref{red:9}.  
\end{observation}

Our goal is to enumerate equivalent matching multicuts by replacing some edges from $M$ with edges from the erased clusters. The branching operations are as follows:

\begin{itemize}
    \item[B1] Let $\{u, v\}$ be a simple edge cluster that is not a matching cluster and was previously erased. Either add edge $uv$ to $M$, or (if possible) add edges $\{u, N(u)\}$ or $\{v, N(v)\}$ to $M$.
    
    \item[B2] Let $C'$ be a matching cluster that has the same neighborhood as some matching cluster $C$ of $\mathcal{C}_{\neq 2}$. Either replace edges $E(C, N(C))$ in $M$ by $E(C', N(C'))$ and remove $C$ from $\mathcal{C}_{\neq 2}$, or continue.
    
    \item[B3] Let $C'=\{u, v\}$ be a matching cluster that has the same neighborhood as another matching cluster $C_1$ of $\mathcal{C}_{2}$. If $C_1$ is painted blue and there is no $C_2 \in \mathcal{C}_{2}$ with the same neighborhood, do nothing. Otherwise, either replace the edges $E(C_1, N(C_1))$ in $M$ with $E(C', N(C'))$ and remove $C_1$ from $\mathcal{C}_{2}$; or add the edge $uv$ to $M$; or continue.
\end{itemize}

After applying steps 1-5 we have a \DFPT\, algorithm parameterized by $dc(G)$ for \pname{Enumeration Matching Multicut}.

%% file: kernel_lb.tex
\section{Kernelization lower bound for distance to cluster}

Since we do have a \DFPT\ algorithm for the vertex-deletion distance to cluster parameterization, it is natural to ask whether we can build a PDE kernel of polynomial size.
In this section, we show this in the negative by presenting an exponential lower bound for \pname{Matching Multicut} under this parameterization.

To obtain our result, we first show a kernelization lower bound for \pname{Set Packing}. In this problem, we are given a ground set $X$, a family $\mathcal{F} \subseteq 2^X$, and an integer $k$, and are asked to find $\mathcal{F}' \subseteq \mathcal{F}$ of size at least $k$ such that for any $A, B \in \mathcal{F}'$ it holds that $A \cap B = \emptyset$.
In particular, we are going to prove Theorem~\ref{thm:kernel_pack_lb}.

\begin{theorem}
    \label{thm:kernel_pack_lb}
    \pname{Set Packing} has no polynomial kernel when parameterized by $|X|$ unless $\NP \subseteq \coNP/\poly$.
\end{theorem}

Our proof is based on an OR-cross-composition~\cite{cross_composition} from \pname{Set Packing} onto itself under the desired parameterization.
To this end, we denote our input collection of \pname{Set Packing} instances by $\{(Y_1, \mathcal{E}_1, r_1), \dots, (Y_t, \mathcal{E}_t, r_t)\}$.
Moreover, we can assume that $Y_i = \{y_1, \dots, y_n\}$ and $r_i = r$ for all $i \in [t]$ and, w.l.o.g, that $t = 2^\tau$ for some $\tau > 0$; the latter can be easily achieved by copying any one instance $2^\tau - t$ times and adding it to the input collection, which at most doubles this set if $\tau$ is the minimum integer such that $2^\tau \geq t$.

\smallskip
\noindent\textbf{Construction.} We construct our $(X, \mathcal{F}, k)$ \pname{Set Packing} as follows. 
Our set $X$ is partitioned into the set of input elements $Y$, index elements $S = \{s_0, s_1, \dots, s_k\}$, and a set of bits $\{b_{i, j}, \overline{b}_{i,j} \mid i \in [\tau], j \in [r]\}$.
For simplicity, we define $\bits_j(a)$ to be the set where $b_{i,j} \in \bits_j(a)$ if and only if the $i$-th bit in the binary representation of $a$ is $1$, otherwise we have that $\overline{b}_{i,j} \in \bits_j(a)$.

The family $\mathcal{F}$ is partitioned in selector sets, which we identify as $\mathcal{T} = \{T_1, \dots, T_t\}$, and packing sets $\mathcal{P}$.
Each $T_a$ is defined as $T_a = \{s_0\} \cup \bigcup_{j \in r} \bits_j(\overline{a})$, where $\overline{a}$ is the (positive) bitwise complement of $a$, i.e. $a + \overline{a} = 2^\tau - 1$.
As to our packing sets, for each input instance $(Y, \mathcal{E}_a, r)$, each $C_i \in \mathcal{E}_a$, and each $j \in [r]$, we add to $\mathcal{P}$ the set $C_{a,i,j} = C_a \cup \bits_j(a) \cup \{s_j\}$.
Finally, we set $k = r + 1$.

Intuitively, packing a set $T_a \in \mathcal{T}$ corresponds to solving one instance $(Y, \mathcal{E}_a, r)$ and, since every $T_a$ has $s_0$, only one of them can be picked.
The way that our $\bits$ sets were distributed, picking $T_a$ automatically excludes all elements in $\mathcal{P}$ corresponding to sets present in another instance $(Y, \mathcal{E}_c, r)$.
Finally, the index elements $S$ are used to ensure that at least one instance set is packed.
The next observation follows immediately from the construction of our instance.

\begin{observation}
    Instance $(X, \mathcal{F}, k)$ is such that $|X| \leq |Y| + (r+1)(1 + \log t)$ and $|C| \leq 1 + r\log t$ for all $C \in \mathcal{F}$
\end{observation}

We now prove that our the described algorithm is, in fact, an OR-cross-composition. The next two lemmas, along with the above observation, immediately imply Theorem~\ref{thm:kernel_pack_lb}.

\begin{lemma}
    If some instance $(Y, \mathcal{E}_a, r)$ admits a solution $\mathcal{E}'$, then there is a solution $\mathcal{F}'$ for $(X, \mathcal{F}, k)$.
\end{lemma}

\begin{proof}
    Suppose w.l.o.g. that $|\mathcal{E}'| = r$, as we can simply drop the surplus elements of $\mathcal{E}'$ until it has the desired size.
    Afterwards, we arbitrarily order or solution as $\sigma = \angled{C_1, C_2, \dots, C_r}$.
    To obtain $\mathcal{F}'$, we proceed as follows: first, pick $T_a$ and add it to $\mathcal{F}'$; then, for $C_i \in \sigma$, we add $C_{a, i, i}$ to $\mathcal{F}'$.
    Note that $C_{a, i, i}$ belongs to $\mathcal{F}$ as, for each $C_{a,i} \in \mathcal{E}_a$ we add $r$ copies of it to $\mathcal{F}$, each containing a different bitset and index element.
    Since $|\mathcal{E}'| = r$, we have that $|\mathcal{F}'| = r + 1 = k$ and, since $\mathcal{E}'$ is a packing of $Y$, it follows that $C_{a, j_1, j_1} \cap C_{a, j_2, j_2} \cap Y = \emptyset$ for all $j_1 < j_2 \in [r]$.
    Moreover, if $i \neq \ell$, it follows that $C_{a, i, \ell} \cap C_{a, \ell, \ell} \subseteq Y$ for any $i, \ell \in [r]$, as the index elements in each packing set are different and so are the bitsets when $i \neq \ell$.
    Consequently, $\mathcal{F}' \setminus \{T_a\}$ is a packing of size $k-1$ of $X$.
    Finally, note that, $T_a \cap C_{a, i, i} = \emptyset$, since $T_a \cap (Y \cup S) = \emptyset$ and $\bits_j(a) \cap \bits_j(\overline{a}) = \emptyset$ since $a$ and $\overline{a}$ have that the XOR of their $j$-th bits is always $1$.
    As such, we conclude that $\mathcal{F}'$ is a packing of size $k$ in $X$.
\end{proof}

\begin{lemma}
    If there is a solution $\mathcal{F}'$ for $(X, \mathcal{F}, k)$, then there is some instance $(Y, \mathcal{E}_a, r)$ that admits a solution.
\end{lemma}

\begin{proof}
    We begin by observing that there must be exactly one selector set $T_a \in \mathcal{F}'$: there cannot be more than one, as $s_0 \in T_a \cap T_c$, and there cannot be none, as there are only $r$ index elements different from $s_0$, every packing set has one index element and we would have to pick $r+1$ of them.
    We claim that every packing set $C_{c, i, j} \in \mathcal{F}'$ must be such that $c = a$.
    Note that $C_{c, i, j} \cap T_a = \emptyset$ if and only if the bitsets present in $C_{c, i, j}$ and $T_a$ are disjoint; moreover two bitsets are disjoint if and only if the corresponding integers $f,g$ satisfy $f = \overline{g}$ and, since $T_a$ contains $\bits_j(\overline{a})$, it holds that $c = \overline{\overline{a}} = a$.
    Finally, since every packing set belongs to the same input instance and $\mathcal{F}'$ is a packing of $X$, it holds that, when restricted to $Y$, the packing sets form a packing in $Y$ of size $k-1 = r$.
    Using only the corresponding sets of $\mathcal{E}_a$, we obtain a solution to $(Y, \mathcal{E}_a, r)$, concluding the proof.
\end{proof}

With Theorem~\ref{thm:kernel_pack_lb} now in hand, proving our desired lower bound is almost trivial with a \PPT\ reduction from our \pname{Set Packing} instance $(X, \mathcal{F}, k)$ parameterized by $|X|$ to a \pname{Matching Multicut} instance $(G, \ell)$ parameterized by the vertex-deletion distance to cluster (we assume that $|X| \geq 3$.
This reduction can be accomplished by constructing the incidence graph between $X$ and $\mathcal{F}$: we add to $G$ a set of vertices corresponding to $X$ and edges so that $G[X]$ becomes a clique; in an abuse of notation, we refer to this set as $X$ as well. 
Then, for each $C_i \in \mathcal{F}$, add a clique on $\max\{|C_i|, 3\}$ vertices and, pick $|C_i|$ vertices of it and add a matching between them and the vertices in $X$ corresponding to the elements of $C_i$.
To conclude the construction, set $\ell = k+1$.
Correctness follows immediately from the fact that each clique added to $G$ is indivisible and at least $\ell - 
 1 = k$ parts of our cut must be equal to cliques originated from the $C_i$'s.
This proves our result of interest for this section.

\begin{theorem}
    When parameterized by the vertex-deletion distance to cluster, size of the maximum clique, and the number of parts of the cut, \pname{Matching Multicut} does not admit a polynomial kernel unless $\NP \subseteq \coNP/\poly$.
\end{theorem}

%% file: final.tex
\section{Final Remarks}

In this paper, we introduced and studied the \pname{Matching Multicut} problem, a generalization of the well known \pname{Matching Cut} problem, where we want to partition a graph $G$ into at least $\ell$ parts so that no vertex has more than one neighbor outside of its own part.
Specifically, we proved that the problem is \NPH\ on subcubic graphs, but admits a quasi-linear kernel when parameterized by $\ell$ on this graph class.
We also showed an $\ell^{\frac{n}{2}}n^{\bigO{1}}$ exact exponential algorithm based on branching for general graphs.
In terms of parameterized complexity, aside from our aforementioned kernel, we give a $2^{\bigO{t \log t}}n^{\bigO{1}}$ time algorithm for graphs of treewidth at most $t$.
Then, we move on to enumeration aspects, presenting polynomial-delay enumeration kernels for the vertex cover and distance to co-cluster parameterizations, the latter of which was an open problem for \pname{Enum Matching Cut}. Finally, we give a \DFPT\ algorithm for the distance to cluster parameterization, and show that no polynomial-sized PDE kernel exists unless $\NP \subseteq \coNP/\poly$. This last result is obtained by showing that \pname{Set Packing} has no polynomial kernel parameterized by the cardinality of the ground set.

For future work, we are interested in further exploring all aspects of this problem, such as graph classes and other structural parameterizations. As with \pname{Matching Cut}, it seems interesting to study optimization and perfect variations of this problem, which may yield significant differences in complexity to \pname{Matching Multicut}.
While \pname{Maximum Matching Multicut} is \NPH\ as \pname{Perfect Matching Cut} is \NPH\ on 3-connected cubic planar bipartite graphs~\cite{barnette_pmc}, the proof does not help in terms of \WHness{1}.
We believe that it in fact is \WH{1}\ parameterized by $\ell$ + number of edges in the cut even on cubic graphs.

Our other questions of interest are mostly in the enumeration realm. In particular, we have no idea if it is possible to enumerate matching cuts on (sub)cubic graphs, and we consider it one of the main open problems in the matching cut literature.
Finally, we are interested in understanding how to rule out the existence of \TFPT\ and \DFPT\ algorithms for a given problem and, ultimately, how to differentiate between problems that admit FPE and PDE kernels of polynomial size and those that do not.